%% file: main.tex
\documentclass[12pt]{article}
\pdfoutput=1

\usepackage{amsmath,amssymb,amsthm,graphicx}
\usepackage{changepage}
\usepackage{enumitem}
\usepackage{algorithm}
\usepackage{algorithmic}
\usepackage[algo2e]{algorithm2e}
\usepackage{scalerel}
\usepackage{accents}
\usepackage{cancel}
\usepackage{xfrac}
\usepackage{subfig}
\usepackage{pgfplots}
\pgfplotsset{compat=newest}
\usetikzlibrary{shapes, shapes.symbols, backgrounds, matrix, calc, arrows, math, arrows.meta, decorations.pathreplacing, decorations.markings, patterns, intersections}
\pgfdeclarelayer{bg}
\pgfsetlayers{bg, main}
\usepackage{tikz}
\usetikzlibrary{patterns}
\usetikzlibrary{intersections}
\usepackage{natbib}
\usepackage{csquotes}
\usepackage{soul}
\usepackage{comment}
\usepackage[hidelinks]{hyperref}
\usepackage[capitalize]{cleveref}

\usepackage[margin=1.25in]{geometry}
\usepackage{setspace}
\theoremstyle{plain}
\newtheorem{theorem}{Theorem}

\usepackage{color-edits}
\addauthor{mf}{blue} 
\addauthor{dp}{red} 
\addauthor{ls}{magenta} 
\addauthor{pd}{orange} 

\usepackage{xspace}

\newcommand*\samethanks[1][\value{footnote}]{\footnotemark[#1]}

\DeclareMathOperator*{\argmax}{argmax}
\DeclareMathOperator*{\argmin}{argmin}

\input{macros}

\usepackage{thm-restate}
\newcounter{1}
\newcounter{2}

\newcounter{4}
\newcounter{5}

\newcounter{8}
\newtheorem{lemma}         [1]{Lemma}
\newtheorem{corollary}     [2]{Corollary}

\newtheorem{definition}    [4]{Definition}
\newtheorem{example}       [5]{Example}

\newtheorem{proposition}        [8]{Proposition}

\title{Ambiguous Contracts\thanks{Accepted for presentation at EC 2023. We thank Tomer Ezra, Uriel Feige  and Yarden Rashti  for insightful feedback, and the co-editor and referees for helpful comments and suggestions.  The work of Feldman and Peretz has been supported by the European Research Council (ERC) under the European Union's Horizon 2020 research and innovation program (grant agreement No. 866132), 
and the NSF-BSF (grant number 2020788).}}

\author{%
Paul D\"utting\thanks{Google Research, Zurich, Switzerland. Email: \texttt{duetting@google.com}}
\and 
Michal Feldman\thanks{%
School of Computer Science, Tel Aviv University, Israel. Email: \texttt{mfeldman@tauex.ac.il, danielperetz@mail.tau.ac.il}}
\and 
Daniel Peretz\samethanks
\and 
Larry Samuelson\thanks{%
Department of Economics, Yale University, New Haven, CT, USA. Email: \texttt{larry.samuelson@yale.edu}}}
\date{September 10, 2024}

\begin{document}

\maketitle

\begin{abstract}
We explore the deliberate infusion of ambiguity into the design of contracts. We show that when the agent is ambiguity-averse and hence chooses an action that maximizes their minimum utility, the principal can strictly gain from using an ambiguous contract, and this gain can be arbitrarily high.  We characterize the structure of optimal ambiguous contracts, showing that ambiguity drives optimal contracts towards simplicity.  We also provide a characterization of ambiguity-proof classes of contracts, where the principal cannot gain by infusing ambiguity. Finally, we show that when the agent can engage in mixed actions, the advantages of ambiguous contracts disappear.
\end{abstract}

\section{Introduction}

\noindent Contracts are often ambiguous.  A construction contract may require that a builder use ``superior materials,'' a  professional services contract may require that a provider exert ``due diligence,'' or ``act as a fiduciary,'' a labor contract may require that the parties ``bargain in good faith,'' and the promotion guidelines of a university may require that a candidate exhibit ``research productivity and excellence.'' 
In each case, the meaning of these phrases may be ambiguous.  

This ambiguity may reflect the impossibility of precise specification.    In contrast, we explore the deliberate infusion of ambiguity as a tool that a principal may employ to increase her contracting power. 

\subsection{The Model}

\noindent We examine a familiar, finite moral hazard problem, augmented to accommodate ambiguous contracts analogously to the treatment of mechanism design problems by \cite{DiTillioEtAl17}.   In the problem we consider, a principal (she) interacts with an agent (he). The agent can take one of $n$ costly actions. Each action $i \in [n]$ induces a probability distribution $p_{i} = (p_1, \ldots, p_m)$ over $m$ outcomes and imposes a non-negative cost $c_i$ on the agent. Each outcome $j \in [m]$ comes with a reward $r_j$ for the principal. The principal cannot directly observe the agent's action, and seeks to influence the agent's choice of action by paying for the stochastic outcomes of the action taken by the agent.  The principal and agent are both risk neutral, and a limited liability constraint forces payments to be non-negative.

A classic contract for this setting includes a payment function $t = (t_1, \ldots, t_m)$, where $t_j$ specifies the non-negative payment from the principal to the agent when outcome $j\in [m]$ is realized. Given a payment function $t$, the agent chooses an action $i\in [n]$ that maximizes his expected payment minus cost. The principal, in turn receives the expected reward of the implemented action, minus the expected payment to the agent under the chosen action.

We are interested in ambiguous contracts.   The source of the ambiguity can be given various interpretations.  It may be that the contract has missing provisions, vaguely worded provisions, or provisions specified in prohibitively copious fine print,%
\footnote{\cite{zuboff2023age} argues that pyramiding cross references can render it impossible to read all of the fine print in a typical contract.}
each case leaving the agent facing a set of possible realized contracts that he cannot refine.  The resolution of the ambiguity may reflect a decision on the part of the principal, the action of a third party such as a court, or random events.

To capture this ambiguity, we define an \emph{ambiguous contract} to be a collection of payment functions $\tau = \{t^1, t^2, \ldots, t^k\}$.  The agent evaluates each action $i \in [n]$ by the minimum utility he could receive from a payment function $t^\ell \in \tau$, and chooses an action  that maximizes this minimum utility. We say that the ambiguous contract implements the selected action.  We impose the consistency condition that,  for the implemented action, every payment function in the support of $\tau$ gives the principal the same payoff.  We motivate this requirement as ensuring that the  principal's ``threat'' that she may  choose any payment function $t \in \tau$ is credible, though we show that the requirement is without loss of generality.

\subsection{Our Contribution}

\noindent Section \ref{sec:model} sets up the model and examines the sets of implementable actions.  We recall the familiar conditions (\cite{HermalinK91}) for an action to be implementable with a classic contract, 
and then characterize implementable actions under ambiguous contracts.  In general, ambiguity expands the set of implementable actions.  

Section \ref{sec:computation} examines optimal ambiguous contracts.  We first show that the principal can use ambiguity to her advantage, either because ambiguity allows her to implement the optimal action under classic contracts at a reduced cost, or because she exploits the ambiguity to implement a different action.  Indeed, a principal wielding ambiguous contracts may optimally induce an action that is impossible to implement under classic contracts.

We then show that ambiguity drives optimal contracts towards simplicity. In general, optimal ambiguous contracts are composed of  at most $\max\{n-1,m\}$ payment functions, each of which can be taken to attach a positive payment to precisely one outcome (i.e., is a single-outcome payment function).  If we restrict attention to principal-agent problems satisfying the monotone likelihood ratio property (MLRP), then optimal ambiguous contracts contain at most two (single-outcome payment) contracts.  If payment functions must be monotone---higher rewards engender higher payments---perhaps for reasons of fairness, regulation, or robustness, then analogous results hold, with payment functions now being step functions (with a single step) rather than paying for only a single outcome.

Section \ref{sec:ambiguity-gap} uses the concept of an \emph{ambiguity gap}---the largest possible ratio of the principal's payoff under an optimal ambiguous contract to that of an optimal classic contract---to quantify how much the principal gains by exploiting ambiguity.  In general, this ratio can be arbitrarily large.  When all rewards are positive, the ambiguity gap is $n-1$, and hence grows arbitrarily large when the number of actions grows large. 

Section \ref{sec:Ambiguity Proofness} defines a class of contracts to be \emph{ambiguity-proof} if it is impossible for the principal to implement an action at a lower expected payment with an ambiguous contract than with a classic contract.  We show that a class of contracts is ambiguity-proof if and only if it is {\em ordered}, in the sense that for any two contracts in the class, one of them attaches a weakly higher payment to every outcome than does the other.  An immediate implication of this result is that the classes of linear contracts is ambiguity-proof, among others.   In contrast, many other natural classes of contracts, such as the classes of  all affine, polynomial or monotone contracts, are \emph{not} ambiguity-proof.

Section \ref{sec:mixing} shows that the advantages of ambiguity disappear if the agent can mix over actions.  The ability to mix provides the agent with more alternative actions, tightening the incentive constraints enough to dissipate any advantage the principal gains from ambiguous contracts. As explained by \cite{Raiffa61} in his assessment of the \cite{Ellsberg61} paradox, mixing allows the agent to transform a situation of ambiguity into one of risk, alleviating the force of ambiguity-aversion.

\subsection{Implications}

\noindent We do not expect to see agents literally facing a bevy of payment functions, wondering which will actually be applied, but we do view our results as helpful in understanding three aspects of real-world contracts.  

First, we believe that contracts typically are ambiguous.  Indeed, it is difficult to imagine a contract that specifies beyond any doubt the implications of every outcome.  The literature has focused on feasibility constraints as the source of such imprecision---it may be prohibitively expensive or impossible to anticipate, describe, or verify the various outcomes.%
\footnote{\cite{AghionHolden11} provides a survey of a literature that has its roots in \citep{Hart88} and \citep{HartMoore88}.}
In contrast, we suggest that ambiguity may be deliberately embraced by the principal as an incentive device.  As our ambiguity gap results show, the gains to the principal from doing so can be large.

Second, a theme that emerges from our results is that optimal ambiguous contracts tend to be simple.  We expect circumstances will often constrain contracts to be monotone.  In this case, the optimal ambiguous contract features a collection of step functions, each with a single step.  In practice, this would take the form of a single contract, specifying bonuses if various performance thresholds are reached, but written sufficiently imprecisely as to make the performance thresholds and bonuses ambiguous.  If the technology satisfies the monotone likelihood ratio property, the number of such performance-dependent bonuses is small, namely two.  We thus have a contract that requires at least adequate performance in order to elicit payment,  with a bonus for superior performance,  written so as to allow some ambiguity as to the precise performance thresholds and payments.  We view actual contracts in many circumstances as fitting this description.  For example, an assistant professor may believe that a reasonable research record engenders promotion to associate professor and a raise, with an exemplary record bringing promotion to full professor and a larger raise, with the  thresholds and the amounts of the raises both ambiguous.

Third, ambiguous contracts can be a burden for agents, either because they are more difficult to evaluate and enforce or because they are a weapon for extracting surplus from agents.  Circumstances may accordingly restrict attention to ambiguity-proof classes of contracts.  Our results show that insisting on ambiguity proofness also drives contracts toward (a different notion of) simplicity.  Ordered contracts, with linear contracts as a leading example, are straightforward to process.  We note that commission contracts are common, and in their simplest form are linear.%
\footnote{A contract may include a base payment plus a commission, making it affine rather than linear.  Section \ref{sec:Ambiguity Proofness} notes that the class of all affine contracts is not ambiguity proof, but there are ambiguity-proof classes of affine contracts.} 

The literature has established other reasons why contracts may be imprecise or simple or linear.  In each case we cannot claim to have provided ``the'' explanation, but our work adds another factor to the list of considerations.

\subsection{Related Literature}

\noindent We work with a familiar hidden-action moral hazard problem, as in \citet{Holmstrom79}, \citet{GrossmanHart83}, and \citet[][Chapter 4]{LaffontM09}, with the friction arising out of limited liability (as in \cite{innes1990limited}) rather than risk aversion.  In contrast to much of the moral hazard literature, our principal offers ambiguous contracts to an ambiguity-averse agent.  We implement the agent's ambiguity aversion by modeling the agent as maximizing his max-min utility \citep{Schmeidler89,GilboaS93}. 

A flourishing literature examines design problems in the face of non-Bayesian uncertainty.  One branch of this literature examines models in which the principal entertains non-Bayesian uncertainty about the agents.  \cite{BergemannSchlag11} examine monopoly pricing on the part of a principal with ambiguous beliefs about buyers' valuations. \cite{CarrascoEtAl2018} examine screening problems in which the principal is only partially informed of the distribution of agent preferences. \cite{Carroll15},   \cite{CarrollW22} and \cite{Kambhampati23} examine moral hazard problems in which the principal has ambiguous beliefs about the set of actions the agent can choose from.
\cite{DaiT22} examine a principal who writes contracts to shape the actions of a team of agents, with the principal holding ambiguous beliefs about the actions available to the agents.
\cite{DuttingRT19} examine moral hazard problems in which the principal has ambiguous beliefs about the distribution of outcomes induced by the agent's actions.  

A second branch of the literature examines settings in which the agent has ambiguous beliefs that the principal can potentially exploit.  
\cite{BeaucheneEtAl19} and \cite{cheng2020ambiguous} examine Bayesian persuasion problems in which the sender exploits the ambiguity aversion of the receiver.  \cite{BodohCreed12} and \cite{DiTillioEtAl17} examine screening problems with agents who have max-min preferences.  \cite{BSRL14} examine mechanism design problems in which agents have max-min preferences.  \cite{lopomo2011knightian} examine moral hazard problems with agents who have Bewley preferences.  \cite{BoseOP06} consider auctions in which the seller and bidders may both be ambiguity averse.
Our paper is distinguished from the two branches above by examining moral hazard problems in which the agent faces ambiguity concerning the payments attached to outcomes.

The paper most closely related to our work is \cite{DiTillioEtAl17}, who conduct a parallel exercise in the context of a screening model, with a seller allocating an object to an ambiguity-averse buyer.     
They find that the seller can gain from offering an ambiguous mechanism, consisting of a set of simple mechanisms, just as our principal can benefit from offering an ambiguous contract.  In each case, the gain comes from using the agent's ambiguity aversion to relax incentive constraints.  \cite{DiTillioEtAl17} show that an optimal ambiguous mechanisms contains at most $N-1$ mechanisms, where $N$ is the number of agent types (as opposed  to the number of actions, in our case), each of which takes a simple form reminiscent of our single-outcome payment functions.  Finally, they show that as the number of agent types grows arbitrarily large, the principal comes arbitrarily close to extracting all of the surplus.

\cite{DiTillioEtAl17} impose a consistency condition on ambiguous mechanisms, analogous to the consistency condition we impose on ambiguous contracts.  
In their case this restriction is substantive rather than sacrificing no generality, reflecting the differing structure of the incentive constraints that arise in screening and moral hazard problems. 
\cite{DiTillioEtAl17} do not have counterparts of our findings that the number of payment functions in an optimal ambiguous contract  is precisely two in the presence of the MLRP condition, though they maintain throughout the screening counterpart of this assumption, in the form of a single-crossing condition.

An implication of our  results is that  in the context of moral hazard problems,  ambiguity and max-min utility drive optimal designs towards simplicity. We thus join a literature, with \cite{holmstrom1987aggregation} as a key early entry, endeavoring to explain why actual contracts in moral hazard settings tend to be simple, in contrast to their theoretical counterparts.   \cite{Carroll15}, \cite{CarrollW22}, and \cite{DaiT22} show that a principal who is uncertain of the actions available to an agent and who has max-min preferences will optimally choose a linear contract.
\cite{DuttingRT19} show that the same holds for a principal who is uncertain about the technology by which actions turn into outcomes and who has max-min preferences.
These papers thus show that ambiguity aversion on the part of the principal can lead to linear contracts, whereas we find linear contracts may be attractive as a device for preventing the principal from exploiting the agent's ambiguity aversion.  In our setting, exploiting the agent's ambiguity aversion leads the principal to an alternative class of simple contracts, consisting of single-outcome payment functions or step functions (when payment functions must be monotone), and including only two such functions when the MLRP condition holds.

\section{The Model}\label{sec:model}

Our starting point is the classic hidden-action principal-agent model of contract theory  (e.g., \citet{Holmstrom79}, \citet{GrossmanHart83}, and \citet[][Chapter 4]{LaffontM09}).
A principal (she) induces an agent (he) to take a costly, unobservable action by writing a contract that attaches payments to the observable, stochastic outcomes of the chosen action.
We augment the model by introducing the notion of an \emph{ambiguous contract} and examining optimal ambiguous contracts.

\subsection{The Principal-Agent Model}

The basic ingredients of the principal-agent model apply to both classic and ambiguous contracts.

\begin{definition}[Instance] An instance $(c,r,p)$ of the principal-agent problem with $n$ actions and $m$ outcomes is specified by: 
\begin{itemize}
\item For each action $i \in [n]$ a non-negative cost $c_i \in \mathbb{R}_+$. We write $c = (c_1, \ldots, c_n)$ for the vector of costs, and sort the actions so that $c_1 \leq c_2 \leq \ldots \leq c_n$.
\item For each outcome $j \in [m]$, a  \emph{reward} $r_j \in \mathbb{R}$. We write $r = (r_1, \ldots, r_m)$ for the vector of rewards, and sort outcomes so that $r_1 \leq r_2 \leq \ldots \leq r_m$.
\item For each action $i \in [n]$, a probability distribution $p_i \in \Delta^m$. We use $p_{ij}$ to denote the probability of outcome $j$ under action $i$. 
\end{itemize}
\end{definition}

We use $\Reward_i=\sum_{j=1}^{m}p_{ij} r_j$ to denote the expected reward of action $i$, and write $\Welfare_i = \Reward_i - c_i$ for action $i$'s \emph{expected welfare}.

The agent retains the option of not participating.   To capture this, we assume throughout that action $1$  
is a zero-cost action, 
that leads with probability $1$ to an outcome that we interpret as the status-quo outcome. 
As we explain in Sections \ref{casper}--\ref{amber}, this does not limit the generality of the model, but leads to a unified treatment of the incentive compatibility and individual rationality constraints.  

The literature often focuses on instances that satisfy the monotone likelihood ratio property:

\begin{definition}[MLRP]\label{def:MLRP}
An instance \instance satisfies the {\em MLRP} (monotone likelihood ratio property) if for any two actions $i,i'$ such that $c_i > c_{i'}$, it holds for all $j'>j$ that:
\[
p_{ij'}p_{i'j}\ge p_{ij}p_{i'j'}.
\]
\end{definition}

Intuitively, the MLRP condition requires that more costly actions are more likely to yield high outcomes.

The specification \instance is known to both the principal and the agent.  The agent's action is known only by the agent, while realized outcomes are observed by both the principal and agent.

A payment function  $t : [m]\rightarrow \mathbb R_+$ identifies a payment made by the principal to the agent upon the realization of each outcome, with the payment $t(j)$ made in response to outcome $j$ typically denoted by $t_j$.  Payments are attached to outcomes rather than actions because the principal can observe only the former.  
We assume that payments are non-negative.  This is a standard  limited liability assumption.

In many cases, considerations of fairness, regulation, 
or robustness may restrict attention to monotone payment functions.

\begin{definition}
The payment function $t$ is {\em monotone} if outcomes generating larger rewards engender  (at least weakly) larger payments:  $r_j\ge r_{j'}\implies t_j\ge t_{j'}$ for all $j,j'\in [m]$.
\end{definition}

\subsection{Classic Contracts}\label{casper}

We first describe the classic setting, in which a contract,  denoted by $\contract$, is a payment function $t$  and a recommended action $i\in [n]$.  The interpretation is that the principal posts a contract, the agent observes the contract and chooses an action and bears the attendant cost, an outcome is drawn from the distribution over outcomes induced by that action, and the principal  makes the payment to the agent specified by the contract.   The inclusion of a recommended action in the contract allows us to capture the common presumption that the agent ``breaks ties in favor of the principal''.  

More precisely, an agent who chooses action $i'$ when facing a contract $\contract$ garners expected utility
\[
\UAmid{i'}{t}~~=~~ \sum_{j=1}^mp_{i'j} t_j-c_{i'} ~~=~~ \Payment{i'}{t}-c_{i'},
\]
given by the difference between the expected payment $\Payment{i'}{t}$ and the cost $c_{i'}$.  The resulting principal utility is $\UPmid{i'}{t} = \Reward_{i'} - T_{i'}(t).$

\begin{definition}[IC contract] 
A contract $\contract$ is incentive compatible (IC) if 
\[
i\in \argmax_{i'\in [n]}  \UAmid{i'}{t},
\]
in which case we say that contract $\contract$ implements action $i$.
\end{definition}

Because payments are non-negative and action $1$ has zero cost, incentive compatibility ensures that the agent secures an expected utility of at least zero, and hence implies individual rationality.

We assume the agent follows the recommendation of an incentive-compatible contract.  If the principal posts the incentive compatible contract $\contract$, the payoffs to the principal and agent are then 
\begin{eqnarray*}
  \UP{\contract}~=~\UPmid{i}{t} &=& \Reward_i-\Payment{i}{t}\\
  \UA{\contract}~=~\UAmid{i}{t}&=& \Payment{i}{t}-c_i.
\end{eqnarray*}

It is without loss of generality to restrict the principal to incentive compatible contracts.  The idea is that an agent facing contract $\contract$  will choose an action that maximizes her expected utility given $t$, and hence the principal might as well name such an action in the contract.  The optimal classic contract implementing each action $i$ can be identified by solving a linear programming problem, presented in Figure \ref{fig:minpaylp} in Appendix \ref{app:implementable}.
The optimal (incentive compatible) classic contract will inevitably induce indifference on the part of the agent, as some incentive constraint will bind.

\subsection{Ambiguous Contracts}\label{amber}

An \emph{ambiguous} contract $\ambcontract = \ambcontractfull$ is a set of
payment functions and a recommended action.  
 If $t \in \{t^1, \ldots, t^k\}$, then we say that $t$ is in the support of $\tau$.

The principal now posts an ambiguous contract, the agent observes the ambiguous contract and chooses an action and bears the attendant cost, a payment function is selected from the support of the ambiguous contract, an outcome is  drawn from the distribution over outcomes induced by that action, and the principal makes the payment to the agent specified by the selected contract.

The agent is a max-min expected utility maximizer \citep{Schmeidler89,GilboaS93}, and so evaluates each action $i$ according to the payment function that minimizes the expected payment of the action.

Formally, given an ambiguous contract $\ambcontract$, the agent's utility for action $i' \in [n]$ is 
\[
\UAmid{i'}{\tau} = \min_{t \in \tau} \UAmid{i}{t}. 
\]
 
\begin{definition}[IC ambiguous contract]
\label{def:ic-amb}
An ambiguous contract $\ambcontract$ is incentive compatible (IC) if
\[
i\in \argmax_{i' \in [n]} \UAmid{i'}{\tau},
\]
in which case we say that ambiguous contract $\ambcontract$ implements action $i$.
\end{definition}
As in the case of a classic contract,  the incentive compatibility constraint implies individual rationality.
It is again without loss to restrict the principal to incentive compatible ambiguous contracts.  

If the principal's expected utility $\UPmid{i} {t}$ under payment scheme $t \in \tau$ is strictly higher than $\UPmid{i}{t'}$ for some $t' \in \tau$, then the principal's ``threat'' that she may use any contract in $\tau$ may not be credible---an agent facing ambiguous contract $\ambcontract$ may fear the principal will contrive to invariably select the payment function $t$ rather than $t'$.  As in \cite{DiTillioEtAl17}, we accordingly  restrict the principal to consistent ambiguous contracts.  

\begin{definition}[Consistency]\label{def:consistent}
An ambiguous contract $\ambcontract = \ambcontractfull$ is {\em consistent} if for any $\ell, \ell' \in [k]$, 
\begin{align}
\label{basic-consistency-new}
\UPmid{i}{t^{\ell}} = \UPmid{i}{t^{\ell'}}.
\end{align} 
\end{definition}

If the principal posts the consistent, incentive compatible ambiguous contract $\ambcontract = \ambcontractfull$, then the induced payment $T_i(\tau)$ can be defined as
\begin{eqnarray*}
\Payment{i}{\tau} =\Payment{i}{t^{\ell}} = \Payment{i}{t^{\ell'}} &&~~~~\forall \ell,\ell' \in [k], \quad\text{and~hence}\\
\UAmid{i}{t^\ell} = \UAmid{i}{t^{\ell'}}&&~~~~\forall \ell,\ell' \in [k],
\end{eqnarray*}
with the first directly implied by consistency and the second following from $U_A(i \mid t) = \Payment{i}{t} - c_i$. The expected utilities $\UP{\langle \tau,i \rangle}$ and $\UA{\langle \tau,i \rangle}$ of the principal and agent are then given by, for any $t\in \tau$, 
\begin{eqnarray*}
U_P(\ambcontract) &=& \UPmid{i}{t} ~=~ \Reward_i-\Payment{i}{\tau}, \quad\text{and}\\
U_A(\ambcontract) &=&  \UAmid{i}{t} ~=~ \Payment{i}{\tau}-c_i.
\end{eqnarray*}

It is without loss of generality to restrict the principal to consistent ambiguous contracts: 

\begin{lemma}
Suppose $\ambcontract$ is incentive compatible.  Then there exists a consistent, incentive compatible ambiguous contract $\ambcontractprime$ from which the principal obtains expected payoff at least $\max_{t \in \tau} \UPmid{i}{t}$.
\end{lemma}

\begin{proof}
Consider an incentive compatible ambiguous contract $\ambcontract$ that is not consistent.
Let the payment functions in $\tau$ be numbered so that 
\[
\UPmid{i}{t^1}= \max_{t\in \tau} \UPmid{i}{t}.  
\]
Suppose 
\[
\UPmid{i}{t^1} > \UPmid{i}{t^2}.
\]
Then it must be that 
\[
\sum_{j=1}^mp_{ij}t^1_j<\sum_{j=1}^mp_{ij}t^2_j.
\]
Let $\theta\in[0,1)$ satisfy
\[
\sum_{j=1}^mp_{ij} \theta t^2_j=\sum_{j=1}^mp_{ij}t^1_j,
\]
and consider the ambiguous contract $\ambcontractprime = \langle\{ t^1, \theta t^2,\ldots, t^k\},i\rangle$, constructed from $\ambcontract$ by replacing payment function $t^2$ with $\theta t^2$.  We have $\UPmid{i}{t^1} = \UPmid{i}{\theta t^2} > \UPmid{i}{t^2}$, and so the principal's payoff is at least as high under $\ambcontractprime$ as under $\ambcontract$.  In addition, we have
\begin{eqnarray*}
\UAmid{i}{\theta t^2} &= &\UAmid{i}{t^1}~~~~~~<~~ \UAmid{i}{t^2} \qquad\text{and}\\
\UAmid{i'}{\theta t^2} &=& \theta \Payment{i'}{t^2} - c_{i'} ~~\le~~\UAmid{i'}{t^2},
\end{eqnarray*}
which imply
\begin{eqnarray*}
\min_{t\in \tau'}\UAmid{i}{t} &=&\min_{t\in \tau}\UAmid{i}{t} ~~~\quad\text{and}\\
\min_{t\in \tau'}\UAmid{i'}{t} &\le&\min_{t\in \tau}\UAmid{i'}{t}~~~~~\forall i'\in[n],
\end{eqnarray*}
which establishes that $\ambcontractprime$ is incentive compatible.  Applying a similar argument to payment functions $t^3,\ldots, t^k$ yields the result.  
\end{proof}

An implication of this result is that we can equivalently view the principal as selecting the contract to be implemented, or as max-min preferences over the outcome of a selection by a third party.

\subsection{Implementability}\label{sec:implementability-characterization}

This section characterizes the actions that are implementable with classic and ambiguous contracts.
The following result for classic contracts is standard (e.g., \citet[Proposition 2]{HermalinK91}). 

\begin{proposition}
\label{prop:implementable}
Action $i \in [n]$ is implementable with a classic contract if and only if there does not exist a convex combination $\lambda_{i'} \in [0,1]$ of the actions $i' \neq i$ that yields the same distribution over rewards $\sum_{i' \neq i}\lambda_{i'} p_{i'j} = p_{ij}$ for all $j$ but at a strictly lower cost $\sum_{i'} \lambda_{i'} c_{i'} < c_i$.
\end{proposition}

For completeness, we provide a proof in Appendix~\ref{app:implementable}. 
In contrast, the conditions for implementing an action with ambiguous contracts are more permissive:

\begin{proposition}
\label{hadyn}
Action $i \in [n]$ is implementable with an ambiguous contract if and only if there is no other action $i' \neq i$ such that $p_{i'} = p_{i}$ and $c_{i'} < c_i$. 
\end{proposition}

\begin{proof} 
We first show that if there exists an action $i' \neq i$ such that $p_{i'} = p_{i}$ and $c_{i'} < c_{i}$, then it is impossible to implement action $i$ with an ambiguous contract. For the sake of contradiction, suppose that ambiguous contract $\ambcontract = \ambcontractfull$ implements action $i$. 
In this case, since $p_{i} = p_{i'}$, we have $\Payment{i}{t^\ell} = \Payment{i'}{t^{\ell'}}$ for all $\ell,\ell' \in [k]$. But then 
$\UAmid{i'}{\tau} = \min_{\ell \in [k]} \Payment{i'}{t^\ell} - c_{i'} > \min_{\ell \in [k]} \Payment{i}{t^\ell} - c_{i} = \UAmid{i}{\tau}$,
contradicting the fact that $\ambcontract$ is incentive compatible. 

Next we show that if there is no action $i' \neq i$ such that $p_{i'} = p_{i}$ and $c_{i'} < c_{i}$, then action $i$ can be implemented with an ambiguous contract. In this case, 
for each action $i' \neq i$, either (i) $p_{i'} \neq p_{i}$ or (ii) $p_{i'} = p_{i}$ and $c_{i'} \geq c_i$. Let $A$ be the actions of type (i). 
If $A$ is empty, then $i$ must be a zero-cost action.
A (consistent) ambiguous contract for implementing that action is $\langle \tau, i \rangle$ with $\tau = \{(0,\ldots,0)\}$. 

Assume $A$ is nonempty.  We construct  an ambiguous contract $\langle\tau,i\rangle$ for implementing action $i$ that has one contract $t^{i'}$ for each action $i' \neq i$ of type (i).
For each action $i' \in A$, let $j(i')$ be an outcome $j$ such that $p_{ij}/p_{i'j}$ is maximal. Note that $p_{ij(i')}/p_{i'j(i')} > 1$. Let 
\[
T = \max_{i' \in A} \left\{\min \left\{x \geq 0 \;\middle|\; p_{i j(i')} \cdot \frac{x}{p_{ij(i')}} - c_i \geq p_{i'j(i')} \cdot \frac{x}{p_{ij(i')}} - c_{i'} \right\}\right\}.
\] 
For each $i' \in A$, let $t^{i'}_{j(i')} = T/p_{ij(i')}$ and $t^{i'}_{j'} = 0$ for $j' \neq j(i')$. 

We conclude by verifying that $\ambcontract = \langle \{t^{i'} \mid i ' \in A\}, i \rangle$ is a (consistent) ambiguous contract that implements action $i$.
It is easy to check consistency.

To see that $\langle \tau, i \rangle$ is incentive compatible, first consider actions $i' \neq i$ of type (ii). For these actions we have
\[
\UAmid{i'}{\tau} = \UAmid{i}{\tau} + c_{i} - c_{i'} \leq \UAmid{i}{\tau},
\]
where we used that $p_{i'} = p_{i}$ and $c_{i'} \geq c_{i}$.

Next consider actions $i' \neq i$ of type (i). For these actions, there must be a $T_{i'} \geq 0$ with $T_{i'} \leq T$ such that
\[
p_{i j(i')} \cdot \frac{T_{i'}}{p_{ij(i')}} - c_i \geq p_{i'j(i')} \cdot \frac{T_{i'}}{p_{ij(i')}} - c_{i'}.
\]
Since $T_{i'} \leq T$ and $p_{ij(i')} > p_{i'j(i')}$ this implies 
\begin{align*}
\UAmid{i}{\tau} = p_{i j(i')} \cdot \frac{T}{p_{ij(i')}} - c_i 
&\geq p_{i'j(i')} \cdot \frac{T}{p_{ij(i')}} - c_{i'}\\
&= \min_{i'' \in A}\left( p_{i'j(i'')} \cdot \frac{T}{p_{ij(i'')}}- c_{i'}\right) = \UAmid{i'}{\tau},
\end{align*}
where the first equality holds by consistency, the second equality holds by definition of $j(i')$, and the final equality holds by definition.
\end{proof}

The following example presents an action that is implementable with an ambiguous contract but not a classic contract.

\begin{example}[Action implementable with ambiguous but not classic contract] 
\label{mozart}
{\em Consider the instance  in Figure \ref{mahler}.
\begin{figure}[t]
\begin{center}
\begin{tabular}{|c|ccc|c|}
\hline
\text{rewards:}& $r_1 = 0$ & $r_2 = 2$ & $r_3 = 2$  & \text{costs} \\[0ex] \hline
action $1$: &$1$&$0$&$0$&$c_1=0$\\
action $2$: &$0$& $1$ & $0$ & $c_2 = 1$\\
action $3$: &$0$& $0$ & $1$ & $c_3 = 1$\\
action $4$: &$0$& $1/2$ & $1/2$ & $c_4 = 3$\\[0ex] \hline
\end{tabular}
\end{center}
\caption{Instance \instance for Example \ref{mozart}.}\label{mahler}
\end{figure}
A half/half combination of actions 2 and 3 gives the same distribution over outcomes as action 4, but at a lower cost, and hence no classic contract can implement action 4.  However, the ambiguous contract $ \langle\tau,4\rangle=\langle \{ t^1,t^2\},4 \rangle$ with $t^1=(0,6,0)$ and $t^2=(0,0,6)$ implements action 4 
with the minimum possible expected payment of $\Payment{4}{\tau} = c_4 = 3$.
\hfill\rule{.1in}{.1in}}
\end{example}

\section{Optimal Ambiguous Contracts}
\label{sec:computation}

We first show, in Section~\ref{advil}, that the principal can benefit by employing ambiguous rather than classic contracts.
This observation motivates us to study the structure of optimal ambiguous contracts.
Section \ref{sec:opt-general} establishes that optimal ambiguous contracts taking a particularly simple form always exist. 
Section \ref{sec:opt-monotone} establishes the counterparts of these results for monotone contracts.

\subsection{The Advantage of Optimal Ambiguous Contracts}
\label{advil}

In the following simple variation of Example \ref{mozart}, 
the principal gains from ambiguity by implementing the same action as in the best classic contract, though less expensively.

\begin{example}[Strict improvement] 
\label{ex:improve}
{\em Consider the instance shown in Figure \ref{plowshare}. 
\begin{figure}[t]
\begin{center}
\begin{tabular}{|c|ccc|c|}
\hline
\text{rewards:}& $r_1 = 0$ & $r_2 = 2$ & $r_3 = 2$  & \text{costs} \\[0ex] \hline
action $1$: &$1$&$0$&$0$&$c_1=0$\\
action $2$: &$1/2$& $1/2$ & $0$ & $c_2 = 1/4$\\
action $3$: &$1/2$& $0$ & $1/2$ & $c_3 = 1/4$\\
action $4$: &$0$& $1/2$ & $1/2$ & $c_4 = 3/4$\\[0ex] \hline
\end{tabular}
\end{center}
\caption{Instance $(c,r,p)$ for Example \ref{ex:improve}.\label{plowshare}}
\end{figure}
The best classic contract is $\langle (0,1,1), 4 \rangle$, yielding the principal an expected utility of $1$. Indeed, the best classic contract implementing action $2$ is $\langle(0,1/2,0),2\rangle$, for a principal's utility of $3/4$, and the same holds for action $3$, with the contract $\langle (0,0,1/2), 3 \rangle$.

An optimal ambiguous contract is  $\langle \tau,4\rangle=\langle \{t^1,t^2\},4 \rangle$, with 
$t^1=(0,3/2,0)$ and $t^2=(0,0,3/2)$. 
The worst payment function in $\tau$ for action $2$ is $t^2$, giving the agent an expected payment of $0$.  Similarly, the worst payment function for action $3$ is $t^1$, for an expected payment of $0$.  Thus, both actions $2$ and $3$ give the agent negative utilities.  In contrast, the expected payment for action $4$ is $3/4$ under both $t^1$ and $t^2$, giving the agent an expected utility of $0$.  The ambiguous contract $\langle \tau, 4 \rangle$ thus implements action $4$, with an expected payment of $3/4$, and an expected utility for the principal of $5/4$ strictly higher than her optimal utility under a classic contract.}  
\hfill\rule{.1in}{.1in}
\end{example}

We next show that the same phenomenon can occur in instances satisfying the MLRP condition. In this example, the optimal ambiguous contract implements a different action than the one implemented by the optimal classic contract.

\begin{example}[Strict improvement under MLRP]\label{ex:improve-mlrp}
{\em 
Consider the instance shown in Figure \ref{bottlenose}. 
\begin{figure}[t]
\begin{table}[H]
     \centering
    \scalebox{0.9}{
    \begin{tabular}{|c|c|c|c|c|c|c|}
    \hline
        \rule{0pt}{2ex}  \hspace{1.0mm} rewards: \hspace{1.0mm}  & \hspace{3.0mm} $r_1 = 0$ \hspace{3.0mm} & \hspace{3.0mm} $r_2 = 32$ \hspace{3.0mm} & \hspace{3.0mm} $r_3 = 33$ \hspace{3.0mm} & \hspace{3.0mm} $r_4 = 34$ \hspace{3.0mm} & \hspace{3.0mm} costs \hspace{3.0mm} \\[0ex] \hline
        action $1$: & $1$ & $0$ & $0$ & $0$ & $c_{1} = 0$\hspace*{5pt}\\[0ex]
        action $2$: & $0.4$ & $0.6$ & $0$ & $0$ & $c_{2} = 1$\hspace*{5pt}\\[0ex]
        action $3$: & $0$ & $0.5$ & $0.5$ & $0$ & $c_3 = 11$ \\[0ex] 
        action $4$: & 0 & $0$ & $0.6$ & $0.4$ & $c_4 = 12$ \\[0ex]  \hline
    \end{tabular}}
\end{table}
\caption{Instance $(c,r,p)$ for Example \ref{ex:improve-mlrp}.
\label{bottlenose}}
\end{figure}
In this instance, the principal can implement both action $2$ and action $4$ with a classic contract, with an expected payment equal to the agent's respective cost. Possible contracts that achieve this include 
$\langle (0,5/3,0,0), 2 \rangle$ 
and $\langle  (0,0,0,30),4 \rangle$. 
The resulting principal utility is $R_2 - c_2 = 18.2$ 
for action $2$ and $R_4 - c_4 = 21.4$
for action $4$. The cheapest way to implement action $3$ (e.g., by solving the LP in Figure~\ref{fig:minpaylp} in Appendix \ref{app:implementable}) is via contract
$\langle t = (0,25/12,245/12,0), 3 \rangle$, yielding a utility of $R_3 - \Payment{3}{t} = 21.25.$ 
The maximal utility the principal can achieve with a classic contract is thus $21.4$.
Contrast this with the optimal ambiguous contract $\langle  \{t^1,t^2\},3 \rangle$, consisting of the two payment functions $t^1 = (0,22,0,0)$ and $t^2 = (0,0,22,0)$. This ambiguous contract implements action $3$ with an expected payment equal to the agent's cost, for a principal utility of $R_3 - c_3 = 21.5.$ \hfill\rule{.1in}{.1in}}\end{example}

In our next example, the principal gains from using an ambiguous contract to implement an action that {\em cannot} be implemented with a classic contract.

\begin{example}[Action optimal with ambiguous contract but not implementable with classic contract] 
\label{bach}
{\em Consider the instance shown in Figure \ref{crusell}.
\begin{figure}[t]
\begin{center}
\begin{tabular}{|c|cccc|c|}
\hline
\text{rewards:}& $r_1 = -200$ & $r_2 = 0$ & $r_3 = 21$  & $r_4=21$&\text{costs} \\[0ex] \hline
action $1$: &$0$&$1$&$0$&$0$&$c_1=0$\\
action $2$: &$0.1$& $0$ & $0.9 $&$0$&$c_2=8 $\\
action $3$: &$0.1$& $0$ & $0$&$0.9$ &$c_3 = 8$\\
action $4$: &$0$&$0$& $1$   & $0$ &$c_4 = 10$\\
action $5$: &$0$& $0$ &$0$  & $1$ &$ c_5= 10$\\
action $6$: &$0$&$0$& $0.5$   & $0.5$ & $c_6 = 11$\\[0ex] \hline
\end{tabular}
\end{center}
\caption{Instance $(c,r,p)$ for Example \ref{bach}.
\label{crusell}}
\end{figure}
Action $6$ cannot be implemented by a classic contract, with the half/half combination of actions $4$ and $5$ giving the same distribution over outcomes at a lower cost.  Actions $2$ and $3$ have a negative expected welfare, and so will never be optimal for the principal. 
Actions $4$ and $5$ can both be implemented with a classic contract, and yield the same maximal utility for the principal.
Optimal classic contracts for these actions include $\langle (0,0,20,0), 4 \rangle$ and $\langle (0,0,0,20),5 \rangle$, each giving the principal an expected utility of $1$.  In contrast, the optimal ambiguous contract $\langle\{(0,0,22,0),(0,0,0,22)\},6\rangle$ implements action 6, for an expected utility of 10.
\hfill\rule{.1in}{.1in}}
\end{example}

We conclude with an example showing that an ambiguous contract may benefit {\em both} the principal and the agent. Clearly, this can only happen when the optimal action under an ambiguous contract differs from the optimal action under classic contracts.

\begin{example}[Ambiguous contracts may benefit both principal and agent]\label{hardy} 
{\em Consider the instance shown in Figure \ref{laurel}.
\begin{figure}[t]
\vspace*{10pt}
\begin{center}
\begin{tabular}{|c|ccc|c|}
\hline
\text{rewards:}& $r_1 = 0$ & $r_2 = 9$ & $r_3 = 9$&\text{costs} \\[0ex] \hline
action $1$: &$1$&$0$&$0$&$c_1=0~~$\\
action $2$: &$0.6$&  $0.3 $&$0.1$&$c_2=0.6 $\\
action $3$: &$0.6$&  $0.1$&$0.3$ &$c_3 = 0.6$\\
action $4$: &$0.2$& $0.4$   & $0.4$ & $c_4 = 3~~$\\[0ex] \hline
\end{tabular}
\end{center}
\caption{Instance $(c,r,p)$ for Example \ref{hardy}.}
\label{laurel}
\end{figure}
An optimal classic contract is $\langle(0,2,0), 2\rangle$, implementing action $2$ with utilities $0$ and $3$ 
to the agent and principal.  The ambiguous contract $\langle \{(0,8,0), (0,0,8)\},4\rangle$ implements action $4$ with utilities $0.2$ and $4$ 
to the agent and principal. 
\hfill\rule{.1in}{.1in}}
\end{example}

\subsection{The Structure of Optimal Ambiguous Contracts}
\label{sec:opt-general}

We now investigate the structure of optimal ambiguous contracts.  We first introduce the simplicity notion of a {\em single-outcome payment} contract.

\begin{definition}[SOP payment function]
\label{def:SOP}
A payment function  $t = (t_1,t_2,\ldots, t_m)$ is a \emph{single-outcome payment function} if there exists an outcome $j \in [m]$ such that $t_{j} > 0$, and for any outcome $j' \neq j$, $t_{j'} = 0$.
\end{definition}

\noindent An ambiguous contract is a single-outcome payment {\em contract} if all of its payment functions have this property.  Proposition \ref{hadyn} used SOP contracts to establish the sufficiency of conditions for implementation under ambiguous contracts.

The following theorem shows that it is without loss of generality to consider ambiguous SOP contracts, and that at most $\min\{m,n-1\}$ payment functions are needed.  In Proposition~\ref{prop:sop-tight} in Appendix~\ref{app:sop-tight} we show that this bound is tight.

\begin{theorem}[Optimal ambiguous contracts]
\label{thm:wlogSOP}
For every IC ambiguous contract $\langle \tau, i\rangle$, there exists an IC ambiguous contract $\langle \tau', i \rangle$, containing at most $\min\{m,n-1\}$ payment functions, such that: 
\begin{enumerate}
    \item $T_i(\tau')=T_i(\tau)$ (i.e., both contracts have the same expected payment and hence same expected payoff to the principal).
    \item For every  $t' \in \tau'$, $t'$ is an SOP payment function.
\end{enumerate}
\end{theorem}

\begin{proof}
Let the ambiguous contract $\tau$ implement action $i$. Let $J = \{j \in [m] \mid p_{i j}>0 \}$. For every $j \in J$,  
consider the SOP payment function with payment $\frac{T_{i}(\tau)}{p_{i j}}$ for outcome $j$. Let $\tau'$ be the ambiguous contract consisting of these SOP payment functions.  By construction, $\tau'$ satisfies properties (1)--(2).  We show that $\tau'$ implements $i$.   Consider an action $i'\neq i$.  Because $\tau$ implements $i$, there exists $t \in \tau$ with 
\[
c_{i}-c_{i'}~\le~ T_{i}(\tau) - \sum_{j} t_jp_{i'j} ~=~ \sum_{j} t_j p_{ij}-\sum_{j} t_j p_{i'j}.
\]
To show that $\tau'$ implements $i$, it suffices to show 
\[
c_{i}-c_{i'}~\le~ T_{i}(\tau) - \min_{j \in J} p_{i'j}\frac{T_{i}(\tau)}{p_{i j}}~ =~ \sum_{j} t_j p_{i j}-\min_{j\in J}\frac{p_{i'j}}{p_{i j}}\sum_{j} t_j p_{i j}.
\]
Combining these, it suffices to show
\[
\min_{j \in J}\frac{p_{i'j}}{p_{i j}}\sum_{j} t_j p_{i j}~\le~ \sum_{j} t_j p_{i'j},
\]
which is equivalent to the obvious statement that
\[
\min_{j \in J}\frac{p_{i'j}}{p_{i j}}~\le~ \frac{\sum_{j}t_jp_{i'j}}{\sum_{j}t_jp_{i j}}.
\]
Notice that $\tau'$ consists of at most $m$ SOP payment functions (in fact, at most $|J|$ payment functions).  If $m>n-1$, one can eliminate from $\tau'$ every payment function that does not minimize the expected payoff to one of the alternatives $i'\neq i$,
leaving at most $n-1$ SOP payment functions.
\end{proof}

\noindent An implication of this result is that if the agent has only two feasible actions, then there exists an optimal ambiguous contract with at most $n-1=1$ 
payment functions,  which is a classic contract.  Hence, with only two actions, ambiguous contracts cannot improve on classic contracts for the principal.

With the help of Theorem~\ref{thm:wlogSOP}, we can show that the contract $\langle \tau, i \rangle$ that we constructed to establish the sufficiency of the conditions for implementation (in the proof of Proposition~\ref{hadyn}) is optimal.

\begin{proposition}\label{prop:opt-ambiguous}
    Suppose action $i \in [n]$ is implementable by an ambiguous contract. Let $A = \{i' \neq i \mid p_{i'} \neq p_{i}\}$. If $A = \emptyset$, then $c_i = 0$, and the IC contract $\langle \{(0,\ldots,0)\}, i \rangle$ is optimal for action $i$. Otherwise, for each $i' \in A$ let $j(i')$ be an outcome such that $p_{ij(i')}/p_{i'j(i')}$ is maximal. Let 
    \[
    T = \max_{i' \in A} \left\{\min \left\{x \geq 0 \;\middle|\; p_{i j(i')} \cdot \frac{x}{p_{ij(i')}} - c_i \geq p_{i'j(i')} \cdot \frac{x}{p_{ij(i')}} - c_{i'} \right\}\right\}.
    \] 
    For each $i' \in A$, let $t^{i'}_{j(i')} = T/p_{ij(i')}$ and $t^{i'}_{j'} = 0$ for $j' \neq j(i')$. 
    Then the IC contract $\langle \tau,i \rangle = \langle\{t^{i'} \mid i' \in A\}, i\rangle$ is optimal for action $i$.
\end{proposition}

The proof of Proposition~\ref{prop:opt-ambiguous} in Appendix~\ref{app:opt-ambiguous} relies on arguments similar to those used to establish that in classic contracts, with two actions, it is optimal to pay only for the maximum likelihood-ratio outcome 
(e.g., \citet[][Chapter 4.5.1]{LaffontM09}, \citet[Full version, Proposition 5]{DuttingRT19}).

We next show that optimal ambiguous contracts for instances satisfying the MLRP condition admit an even simpler structure, namely an ambiguous contract composed of only two SOP payment functions.

\begin{theorem}[Optimal ambiguous contracts under MLRP]
\label{theo:unified MLRP SOP}
Let \principalAgentSetting be an instance that satisfies the MLRP condition. 
For every IC ambiguous contract $\langle \tau, i\rangle$, there exists an IC ambiguous contract $\langle\tau',i\rangle = \langle\{t^1,t^k\},i\rangle$, such that: 
\begin{enumerate}
    \item $T_i(\tau')=T_i(\tau)$ (i.e., both contracts have the same expected payment and hence the same expected payoff to the principal).
    \item $t^1$ and $t^k$ are SOP payment functions, where $t^1_{\ell} = \frac{\payment{i}{\tau}}{\prob{i}{\ell}}$ for $\ell = \min\{j \in [m] \mid \prob{i}{j} > 0 \}$, and $t^k_h = \frac{\payment{i}{\tau}}{\prob{i}{h}}$ for $h = \max\{j \in [m] \mid \prob{i}{j} > 0 \}$. 
\end{enumerate}
\end{theorem}

The proof of Theorem~\ref{theo:unified MLRP SOP}, which is deferred to  Appendix~\ref{app:proof-of-unified-mlrp-sop}, combines the structural properties established in Theorem~\ref{thm:wlogSOP} with the MLRP condition to argue that two SOP payment functions, introduced in the proof of Theorem \ref{thm:wlogSOP},  suffice.  Payment function $t^k$  ``defeats'' all actions with smaller costs and payment function  $t^1$ ``defeats'' all actions with higher costs.

The ability to restrict attention to SOP contracts allows us to show that optimal ambiguous contracts are relatively easy to identify.  Appendix~\ref{app:algorithms}
shows that there exists an algorithm capable of computing the optimal ambiguous contract 
in time $O(n^2m)$ (and time $O(n^2+m)$ under the MLRP condition).  The key implications of these results is that computation time increases polynomially rather than exponentially in the size of the instance.

\subsection{Optimal Monotone Ambiguous Contracts}
\label{sec:opt-monotone}

In some scenarios it is desired or even required, for reasons of fairness, robustness, or regulation, to restrict attention to monotone payment functions.  A contract whose payment functions are monotone is a monotone contract. 
The monotonicity requirement rules out SOP contracts. The following is a natural alternative simplicity notion for monotone contracts.

\begin{definition}[Step payment function]
\label{def:step-contract}
A payment function $t = (t_1,t_2,\ldots , t_m)$ is a \emph{step} payment function if there exists an outcome $k \in [m]$ and some $x \geq 0$, such that $t_{j} = 0$ for every outcome $j < k$, and $t_{j} = x$ for every outcome $j \geq k$.
\end{definition}

\noindent  A contract composed of step payment functions is a step contract.

The following theorem shows that it is without loss of generality to consider ambiguous monotone contracts that are composed of step payment functions.
In that sense, step contracts are the  analogue of SOP contracts for monotone contracts.

\begin{theorem}[Optimal monotone ambiguous contracts]
\label{theo: monotone ambiguous contract}
For every IC monotone ambiguous contract $\langle \tau, i\rangle$, there exists an IC monotone ambiguous contract $\langle\tau',i\rangle$ consisting of at most $\min\{m,n-1\}$ contracts, such that: 
\begin{enumerate}
    \item $T_i(\tau')=T_i(\tau)$ (i.e., both contracts have the same expected payment and hence the same payoff to the principal).
    \item Every payment function in $\tau'$ is a step payment function.
\end{enumerate}
\end{theorem}

\noindent  Clearly, the theorem holds if we replace ``monotone ambiguous contract'' with ``step ambiguous contract.''

\begin{proof}[Proof of Theorem~\ref{theo: monotone ambiguous contract}]
We construct $\tau'$ as follows. For every action $i' \neq i$ there must be a monotone payment function $t^{i'} \in \tau$, such that 
$\UAmid{i'}{t^{i'}} \le \UAmid{i}{t^{i'}}= \UAmid{i}{\tau}$.
By the same arguments as in the proof of Theorem~\ref{thm:wlogSOP} it is now sufficient to show that there is a step payment function $\hat{t}^{i'}$ to put into $\tau'$ such that (i) $\payment{i}{\tHat^{i'}} = \payment{i}{t^{i'}}$ and (ii) $\payment{i'}{\tHat^{i'}} \le \payment{i'}{t^{i'}}$.  

For actions $i' \neq i$ such that there exists an outcome $\jHat \in [m]$ for which $\sum_{\ell = \jHat}^m \prob{i}{\ell} > 0$ and $\sum_{\ell = \jHat}^m \prob{i'}{\ell} = 0$, set $\tHat^{i'}_{j} = \frac{\payment{i}{t^{i'}}}{\sum_{\ell = \jHat}^m \prob{i}{\ell}}$ for all $j \geq \jHat$ and $\tHat^{i'}_{j}=0$ for all $j < \jHat$.
To see that Condition (i) is satisfied, observe that
\begin{eqnarray*} 
\payment{i}{\tHat^{i'}}  = \sum_{j=1}^m{\prob{i}{j}\cdot \tHat^{i'}_j} =  \sum_{j=\jHat}^m{\prob{i}{j}\cdot \frac{\payment{i}{t^{i'}}}{\sum_{\ell = \jHat}^m \prob{i}{\ell}}} = \payment{i}{t^{i'}} \cdot \frac{\sum_{j=\jHat}^m{\prob{i}{j}}}{\sum_{\ell = \jHat}^m{\prob{i}{\ell}}} = \payment{i}{t^{i'}}.
\end{eqnarray*}
Condition (ii) is satisfied, because $\sum_{\ell = \jHat}^m \prob{i'}{\ell} = 0$ and $\hat{t}_j =0$ for $j < \jHat$ imply that $\payment{i'}{\tHat^{i'}} = 0$, while $\payment{i'}{t^{i'}} \geq 0$ by limited liability.

Consider next actions $i' \neq i$ such that for all $\jHat \in [m]$ where $\sum_{\ell = \jHat}^m{\prob{i}{\ell}} > 0$ it holds that $\sum_{\ell = \jHat}^m{\prob{i'}{\ell}} > 0$. Let 
$$\jHat \in \argmax_{j' \in [m]: \sum_{\ell = j'}^m{\prob{i}{\ell}} > 0}\frac{\sum_{\ell = j'}^m{\prob{i}{\ell}}}{\sum_{\ell = j'}^m{\prob{i'}{\ell}}}.$$ 
Define $\tHat^i$ as follows:
Let 
$\tHat^{i'}_{j} = \frac{\payment{i}{t^{i'}}}{\sum_{\ell = \jHat}^m \prob{i}{\ell}}$ 
for all $j \geq \jHat$, and $\tHat^{i'}_j = 0$ for all $j < \jHat$.
It is again easy to verify that Condition (i) holds. Condition (ii) follows from noting that
\begin{eqnarray*}
T_{i'}(\hat t^{i'})
&=&
\sum_{j=\hat j}^mp_{i'j}\frac{\payment{i}{t^{i'}}}{\sum_{\ell = \jHat}^m \prob{i}{\ell}}
=
\frac{\sum_{j=\hat j}^m p_{i'j}}{\sum_{\ell = \jHat}^m \prob{i}{\ell}}\sum_{j=1}^mp_{ij}t^{'}i_j\\
&=&
\min_{j' \in [m]: \sum_{\ell = j'}^m{\prob{i}{\ell}} > 0}\frac{\sum_{j= j'}^mp_{'j}}{\sum_{j = j'}^m \prob{i}{\ell}}\sum_{j=1}^mp_{i,j}t^{i'}_j
\le
\sum_{j=1}^mp_{ij}t^{i'}_j
= 
T_{i'}(t^{i'}),
\end{eqnarray*}
where the inequality follows by observing that one possible choice for $j'$ is the smallest index such that $p_{ij'} > 0$. 

We still have to show that $|\tau'| \leq \min\{m,n-1\}$.
Clearly, $|\tau'| \leq n-1$ by the construction of $\tau'$ (we add one payment function for every $i' \neq i$).
The fact that $|\tau'| \leq m$ follows by the consistency of $\tau'$, combined with the fact that any payment function $t \in \tau'$ is a step payment function, i.e., $\payment{i}{\tau}$ combined with the outcome in which the step of a payment function occurs, uniquely defines the payment function. 
\end{proof}

We next show that optimal monotone ambiguous contracts for instances satisfying the MLRP condition admit an even simpler structure. Namely, similar to the unrestricted case where we did not impose monotonicity, the MLRP condition implies that an optimal ambiguous contract consists of only two payment functions.

\begin{theorem}[Optimal monotone ambiguous contracts under MLRP]
\label{theo:unified MLRP monotone ambiguous contract}
Let \principalAgentSetting be an instance that satisfies the MLRP condition. 
For every IC monotone ambiguous contract $\langle \tau, i\rangle$, there exists an IC monotone ambiguous contract $\langle\tau',i\rangle=\langle\{t^1,t^k\},i\rangle$  such that: 
\begin{enumerate}
    \item $T_i(\tau')=T_i(\tau)$ (i.e., both contracts have the same expected payment and hence the same payoff to the principal).
    \item $t^1$ and $t^k$ are step payment functions: $\hspace{2.0mm} t^{1}_j = \payment{i}{\tau}$ for all $j \geq \ell$, where $\ell = \min\{j \in [m] \mid \prob{i}{j} > 0 \}$, and $t^k_j = \frac{\payment{i}{\tau}}{\prob{i}{h}}$ for all  $j \geq h$, where $h = \max\{j \in [m] \mid \prob{i}{j} > 0 \}$.
\end{enumerate}
\end{theorem}

The proof of Theorem~\ref{theo:unified MLRP monotone ambiguous contract} appears in Appendix~\ref{app:montone-with-mrlp}, and proceeds by verifying that the contract stated in the second 
bullet satisfies the condition in the first bullet and implements action $i$. 
As in the case of Theorem~\ref{theo:unified MLRP SOP}, $t^k$ protects against all actions with lower cost and $t^1$ protects against all actions with higher cost.

\begin{example}[Strict improvement under MLRP with monotone contracts]
{\em \label{james}
We return to the instance shown in Figure \ref{bottlenose}, but now require contracts to be monotone.  The payment functions for the optimal classic monotone contracts implementing each of the various actions, and the attendant payoffs for the principal, are: 
\begin{eqnarray*}
{\rm action~}1:~~~~~~~~~~~(0,0,0,0)&~~~&~~~~0\\
{\rm action~}2:~~(0,5/3,5/3,5/3)&&18.2\\
{\rm action~}3:~~~\hspace*{2pt}\;~~(0,25,25,25)&&~\;7.5\\
{\rm action~}4:~~~\hspace*{2pt}\;~~~~~(0,0,0,30)&&21.4.
\end{eqnarray*}
An ambiguous contract allows us to improve on the cost of implementing (only) action 3.  An optimal ambiguous contract implements action 3 with the two step payment functions 
\begin{eqnarray*}
t^1&=&(0,11,11,11),\qquad\text{and}\\
t^k&=&(0,0,22,22).
\end{eqnarray*}
for a payoff to the principal of $21.5$.
\hfill\rule{.1in}{.1in}}\end{example}

\section{The Ambiguity Gap}
\label{sec:ambiguity-gap}

Section \ref{advil} confirmed that it can be advantageous for the principal to offer ambiguous contracts.  
To quantify the extent of the potential gains, we introduce the notion of the {\em ambiguity gap}, defined as the worst-case ratio between the principal's utility with and without ambiguity.

We restrict attention throughout this section to instances for which the optimal classic contract induces a non-negative utility for the principal.  Let $\mathcal C(c,r,p)$ and $\mathcal A(c,r,p)$ be the sets of incentive compatible classic contracts and incentive compatible ambiguous contracts, for an instance $(c,r,p)$.

\begin{definition}[Ambiguity gap]
The ambiguity gap $\rho(c,r,p)$ of a given instance $(c,r,p)$ and the ambiguity gap $\rho(\mathcal{I})$ of a class of instances $\mathcal{I}$, are 
\begin{align*}
      \rho(c,r,p) = \frac{\max_{\langle \tau,i\rangle \in\mathcal A(c,r,p)}U_P(\langle \tau, i\rangle)}{\max_{\langle t,i\rangle\in\mathcal C(c,r,p)}U_P(\langle t, i\rangle)}
      \quad\quad \text{and} \quad\quad \rho(\mathcal{I}) = \sup_{(c,r,p) \in \mathcal{I}} \rho(c,r,p).
\end{align*}  

\end{definition}

\subsection{Unbounded Ambiguity Gap in General}
\label{sec:ambiguity-negative}

The following example shows that the ambiguity gap can be arbitrarily large.

\begin{figure}[t]
\begin{table}[H]
    \centering
    \scalebox{0.9}{  
    \begin{tabular}{|c|c|c|c|c|c|c|}
    \hline
        \rule{0pt}{2ex}  \hspace{1.0mm} rewards: \hspace{1.0mm}  & \hspace{3.0mm} $r_1 = -r$ \hspace{3.0mm} & \hspace{3.0mm} $r_2 = -r$ \hspace{3.0mm} & \hspace{3.0mm} $r_3 = 0$ \hspace{3.0mm} & \hspace{3.0mm} $r_4 = r$ \hspace{3.0mm} & \hspace{3.0mm} costs \hspace{3.0mm} \\[0ex] \hline
        action $1$: & $0$  & $0$  & $1$    & $0$    & $c_{1} = 0$\hspace*{5pt}\\[0ex]
        action $2$: & $0.5$ & $0$  & $0$    &$0.5$   & $c_{2} = 10$\hspace*{5pt}\\[0ex]
        action $3$: & $0 $ & $0.5$ & $0$    & $0.5$  & $c_{3} = 10$\hspace*{5pt} \\[0ex] 
        action $4$: & $0.2$ & $0.2$ & $0$    & $0.6$   & $c_4 = 20$\hspace*{5pt}\\[0ex]  \hline
\end{tabular}
} 
\end{table}  
\caption{Instance $(c,r,p)$ for Example~\ref{ex:unbounded-gap}} 
\label{fig:unbounded-gap}
\end{figure}

\begin{example}[Unbounded gap with negative rewards]\label{ex:unbounded-gap}
{\em
Consider the instance shown in Figure \ref{fig:unbounded-gap}, where $r\ge 0$ is a parameter we will allow to vary.  Actions 2 and 3 generate negative welfare, and hence only action 4 is capable (depending on $r$) of producing positive welfare.  Welfare is given by
\[
\max\{0,0.2r-20\},
\]
and is positive if and only if $r>100$.  An optimal classic contract implementing action 4 is $\langle t,4\rangle=\langle(0,0,0,100),4\rangle$, which gives 
\[
U_P(\langle t,4\rangle) 
=0.2r-60,
\]
which is positive if and only if $r>300$.  An optimal ambiguous contract implementing action 4 is $\langle \tau,4\rangle = \langle\{t^1,t^2\},4 \rangle=\langle\{(100,0,0,0),(0,100,0,0)\},4\rangle$, giving
\[
U_P(\langle \tau,4\rangle)
=0.2r-20,
\]
which is positive if and only if $r>100$.  Hence for $r\in (100,300]$, the best classic contract generates a payoff of 0, while the best ambiguous contract generates a positive payoff, yielding an infinite ambiguity gap.}
\hfill\rule{.1in}{.1in}
\end{example}

\subsection{Tight Ambiguity Gap under Non-Negative Rewards}
\label{sec:ambiguity-non-negative}

In contrast to the unbounded ambiguity gap in general instances, we next show that for instances in which all rewards are non-negative  
the ambiguity gap is at most $n-1$ and this is tight.

\begin{proposition}\label{prop:gap-general}
Fix $n \geq 2$.
Let $\mathcal{I}^+_n$ denote the class of all instances with $n$ actions and non-negative rewards. 
The ambiguity gap of $\mathcal{I}^+_n$ is
\[
\rho(\mathcal{I}^+_n) = n-1.
\]
\end{proposition}

\begin{figure}[t]
\begin{center}
\scalebox{0.95}{
\begin{tabular}{|c|ccc|c|}
\hline
\text{rewards:}& $r_1 = 0$ & $r_2 = 0$ & $r_3 = \frac{1}{\gamma^{n-2}}$  & \text{costs} \\[0ex] \hline
action $1$: &$1$&$0$&$0$&$c_1=0$\\
action $2 \leq i \leq n-1$: &$0$& $1-\gamma^{n-i}$ & $\gamma^{n-i}$ & $c_i = \frac{1}{\gamma^{i-2}}-(i-1)+(i-2)\gamma$\\
action $n$: &$\delta$& $0$ & $1-\delta$ & $c_n = \frac{1}{\gamma^{n-2}}-(n-1)+(n-2)\gamma$ 
\\[0ex] \hline
\end{tabular}
}
\end{center}
\caption{Instance $(c,r,p)$ used in the proof of Proposition~\ref{prop:gap-general}.}
\label{fig:gap-general}
\end{figure}

The upper-bound direction of the argument makes use of the notion of a linear contract, which is used to state the following lemma.

\begin{definition}[Linear contract]\label{def:linear}
    Consider an instance $(c,r,p) \in \mathcal{I}^+_n$. A (classic) contract $\langle t,i \rangle$ is \emph{linear} if $t = (\alpha r_1, \ldots, \alpha r_m)$ for some $\alpha\ge 0$.
\end{definition}

For an instance $(c,r,p) \in \mathcal{I}^+_n$, denote by $\mathcal{L}(c,r,p)$ the set of all incentive compatible linear contracts.  Let $W=\max_{i\in[n]}W_i$ denote the maximum welfare. 

The following lemma shows that in instances in which the status-quo outcome has a reward of zero, the principal can achieve a $1/(n-1)$ fraction of the optimal welfare as utility with a linear contract.

\begin{lemma}\label{lem:linear-vs-welfare}
Consider instance $(c,r,p) \in \mathcal{I}^+_n$ in which action $1$ has a cost of $c_1 =0$ and invariably leads to reward $r_1 = 0$. Then there exists a subset of actions $A \subseteq [n]$ with $|A| \leq n-1$ 
and a scalar $\alpha_i\ge 0$ for each $i\in A$ such that each linear contract $\langle t,i \rangle$ with $t = (\alpha_i r_1, \ldots, \alpha_i r_m)$ is IC and 
\begin{align}
\max_{\langle t,i \rangle \in \mathcal{L}(c,r,p)} U_P(\langle t,i \rangle) 
= \max_{i \in A} \; (1-\alpha_i) R_i \geq \frac{W}{n-1}.
 \label{eq:approx-bound}
\end{align}
\end{lemma}

The proof of Lemma~\ref{lem:linear-vs-welfare} is similar to arguments in \citep{DuttingRT19}. For completeness, we provide a proof of this lemma in Appendix~\ref{app:linear-vs-welfare}. 
We are now ready to prove Proposition~\ref{prop:gap-general}.

\begin{proof}[Proof of Proposition~\ref{prop:gap-general}]
We first show the upper bound on the ambiguity gap. To this end, fix any $n \geq 2$ and any instance $(c,r,p) \in \mathcal{I}_n^+$. Clearly, the maximum utility the principal can achieve with an ambiguous contract satisfies $\max_{\langle \tau,i\rangle \in\mathcal A(c,r,p)}U_P(\langle \tau, i\rangle) \leq W$. Thus, in order to prove the upper bound on the ambiguity gap, it suffices to show that the maximum utility the principal can achieve with a classic contract satisfies $\max_{\langle t,i\rangle\in\mathcal C(c,r,p)}U_P(\langle t, i\rangle) \geq W/(n-1)$. Consider using a contract of the form $\langle t,i \rangle$ with $t = (\alpha (r_1 - r_1), \ldots, \alpha(r_m-r_1))$, and let $(c,r',p)$ be a modified instance in which $r'_j = r_j - r_1$ for all $j \in [m]$. Note that from the agent's perspective applying contract $\langle t,i \rangle$ in the original instance, is equivalent to applying contract $\langle t',i\rangle$ with $t' = (\alpha r'_1, \ldots, r'_m)$ in the modified instance.

Let $R'_i$ and $W'_i$ for $i \in [n]$ denote the expected reward and welfare of action $i$ in the modified instance, and let $W' = \max_{i \in [n]} W'_i$. Note that $R_i = r_1+R'_i$ and $W_i = r_1 + W'_i $ for all $i \in [n]$, and hence also $W = r_1 + W'$. Applying Lemma~\ref{lem:linear-vs-welfare} to the modified instance, we know that there exists a set $A \subseteq [n]$ and scalars $\alpha_i$ for $i \in A$ such that
\begin{align*}
\max_{\langle t,i\rangle\in\mathcal C(c,r,p)}U_P(\langle t, i\rangle) &\geq \max_{i \in A}  \big(r_1 + (1-\alpha_i)R'_i\big)\\
&= r_1 + \max_{i \in A} \big( (1-\alpha_i)R'_i \big) \\
&\geq r_1 + \frac{1}{n-1} W'\\
&\geq \frac{1}{n-1} \big(r_1 + W'\big) = \frac{1}{n-1} W,
\end{align*}
where in the last step we used that $r_1 \geq 0$. This completes the proof of the upper bound on the ambiguity gap.

We next show the lower bound on the ambiguity gap. To this end, we vary a lower bound
construction due to  \citet{DuettingRT20}. For $n = 2$ there is nothing to show, so fix any $n \geq 3$.
Let $\gamma, \epsilon \in (0,1)$ and let $\delta = \epsilon \cdot \gamma^{n-2}$. Consider the parameterized instance $(c,r,p)$ with $n$ actions depicted in Figure~\ref{fig:gap-general}. 
Lemma~\ref{lem:atmostone} in Appendix~\ref{app:atmostone} shows that the maximal utility the principal can achieve in this instance with a classic contract is at most $1.$ The argument proceeds by showing an upper bound of $1$ for each action $i \in [n]$. For actions $i =1,2$ the upper bound is immediate, as the welfare of these actions is $W_1 \leq W_2 \leq 1$. For actions $i \in \{3,\ldots,n\}$ the upper bound can be shown by considering only a subset of the IC constraints.

The proof is completed by observing that with an ambiguous contract the principal can implement action $n$, with an expected payment equal to $c_n$. This is enough to show the claim, as the welfare from that action is $W_n = (n-1) - (n-2)\gamma - \epsilon$, and $W_n \rightarrow n-1$ as $\gamma, \epsilon \rightarrow 0$.
The ambiguous contract that achieves this is $\langle \{t^1,t^2\},n \rangle$ 
with $t^1 = (\frac{c_{n}}{\delta},0,0)$ and $t^2 = (0,0,\frac{c_{n}}{1-\delta})$. It is easy to verify that this contract is consistent, and entails an expected payment of $c_n$ for action $n$. It is IC, because for all actions $i \neq n$ it gives a minimum payment of zero. 
\end{proof}

\section{Ambiguity Proofness}
\label{sec:Ambiguity Proofness}

In this section, we explore which classes of contracts are amenable to improvements via ambiguous contracts. We phrase our results in terms of properties of payment functions.

It simplifies the exposition to restrict attention to  
payment functions with the property  that  two outcomes that induce the same reward also induce the same payment.  We can thus think of a payment function as mapping from 
$\mathbb R$ into $\mathbb R_+$.  Within this setting, a \emph{class of payment functions} $\classOfContracts$ is a set (possibly of infinite size) of payment functions $t: \mathbb{R} \rightarrow \mathbb{R}_+$.

We first give the definition of an ambiguity-proof class of payment functions. 
For a given instance $(c,r,p)$, let $\mathcal{C}_\mathcal{T}(c,r,p)$ denote the set of all incentive compatible classic contracts with payment functions from $\mathcal{T}$. Let $\mathcal{A}_\mathcal{T}(c,r,p)$ be the analogous definition for ambiguous contracts.

\begin{definition}[Ambiguity-proof]
A class of payment functions $\mathcal{T}$ 
is \emph{ambiguity-proof} if for any instance \instance and any action $i \in [n]$ it holds that 
\[
\max_{\tau: \langle \tau, i \rangle \in \mathcal{A}_{\mathcal{T}}(c,r,p)} U_P(\langle \tau,i \rangle) \leq \max_{t: \langle t, i \rangle \in \mathcal{C}_{\mathcal{T}}(c,r,p)} U_P(\langle t,i \rangle),
\]
i.e., the principal cannot gain from implementing any action $i$ with an ambiguous rather than a classic contract.
\end{definition}

\noindent For example, the principal-agent setting in Example~\ref{ex:improve} shows that the contract class of all contracts is not ambiguity-proof.

Our condition for ambiguity-proofness will be the following:

\begin{definition}[Ordered class of payment functions]
\label{definition:proper crossing}
A class of payment functions $\classOfContracts$ is  \emph{ordered} if for any two payment functions $t,t' \in \classOfContracts$ it holds that: 
\begin{align*}
    t(x) \geq t'(x) \quad \text{for all $x \in \mathbb{R}$} \quad
    \text{or} \quad 
    t(x) \leq t'(x) \quad \text{for all $x \in \mathbb{R}$.} 
\end{align*} 
\end{definition}

\begin{theorem}[Ambiguity-proofness characterization]
\label{thm:manipulable-characterization}
A class of payment functions $\classOfContracts$ is ambiguity-proof if and only if it is ordered.
\end{theorem}

\begin{proof}
We first show that an ordered class of payment functions is ambiguity-proof. Suppose $\classOfContracts$ is ordered. Consider a (consistent) incentive compatible ambiguous contract $\langle\tau,i\rangle = \langle\{t^1, \ldots, t^k\},i\rangle \in \mathcal{A}_\mathcal{T}(c,r,p)$ with $k \geq 2$.  
Since $\classOfContracts$ is ordered, there must exist a payment function in $\tau$, say $t^1$, with the property that  $t^1_j \leq t_j$ for all $j \in [m]$ and all $t\in \tau$.  Hence, for every action  $i' \in [n]$, we have 
\begin{align*}
U_A(i'\mid t^1)= \sum_{j=1}^m \prob{i'}{j} t^1_j - c_{i'} 
= \min_{t\in \tau}\sum_{j=1}^m \prob{i'}{j}t_j - c_{i'}
= \min_{t \in \tau}   U_A(i' \mid t)=  U_A(i' \mid \tau), 
\end{align*}
%
%
which implies that the classic contract
$\langle t^1,i\rangle  \in \mathcal{C}_\mathcal{T}(c,r,p)$ is incentive compatible. 
Moreover, by consistency of $\langle \tau, i \rangle$, 
it also holds that
\begin{align*}
U_P(\langle t^1,i \rangle) =
U_P(i\mid t^1)=U_P(i \mid \tau) = U_P(\langle \tau, i \rangle).
\end{align*}
Hence, the classic 
contract $\langle t^1, i \rangle$ implements action $i$ at the same cost as does $\langle \tau, i \rangle$, and so $\classOfContracts$ is ambiguity-proof. 
    
We next show that ambiguity-proofness implies ordering, by proving the contrapositive.
Suppose $\classOfContracts$ violates ordering. Then there exist $t,t' \in \classOfContracts$ and $x_1,x_2 \in \mathbb{R}$ such that $t(x_1) > t'(x_1)$ and $t(x_2) < t'(x_2)$.  
Letting $\delta_1 = t(x_1)-t'(x_1)>0$, $\delta_2 = t(x_2)-t'(x_2)<0$, $q_1=\frac{-\delta_2}{\delta_1-\delta_2}$ and $q_2=\frac{\delta_1}{\delta_1-\delta_2}$, we obtain values $q_1,q_2>0$ with $q_1+q_2=1$
satisfying
\begin{equation}
\label{spooky}
    q_1 t(x_1) + q_2 t(x_2) =  q_1 t'(x_1) + q_2 t'(x_2).
\end{equation}
Let $\kappa = \min_{j =1,2} \min\{t(x_j),t'(x_j)\}$. 
Note that $\kappa \geq 0$ by limited liability.
Now consider the following instance with $2$ outcomes $r_1=x_1$ and $r_2=x_2$ and $3$ actions as follows:
\begin{itemize}
    \item Action $i \in \{1,2\}$: $p_{ii}=1$ and $c_i=\min\{t(r_i),t'(r_i)\} - \kappa$.
    \item Action $3$: for $j\in\{1,2\}$, $p_{3j}=q_j$, and $c_{3} = \sum_{j=1}^2{q_j t(r_j)} - \kappa$.
\end{itemize}
Note that this construction ensures that $c_i \geq 0$ for all $i \in \{1,2,3\}$, and that $c_i = 0$ for some $i \in \{1,2\}$, which we can take to be the default action.

We argue that action $3$ can be implemented by an ambiguous contract and cannot be implemented 
by any classic  
contract. 
We first show that action $3$ can be implemented with the ambiguous contract $\langle \tau, 3 \rangle$ where  $\tau = \{t,t'\}$, 
with an expected payment equal to $c_{3} + \kappa$.
To see that $\langle \tau,3 \rangle$ is consistent, note that 
$$
    \payment{3}{t} = \sum_{j=1}^2{q_j \cdot t(r_j)} \stackrel{\eqref{spooky}}{=} \sum_{j=1}^2{q_j \cdot t'(r_j)} =  \payment{3}{t'}.
$$
Since $t$ and $t'$ have the same expected payment for action $3$, the agent's expected utility for taking action $3$ under $t$ and $t'$ is the same, and equals
$$
    \UAmid{3}{t} = \UAmid{3}{t'} = \sum_{j=1}^2{q_j \cdot t(r_j)}- c_{3} = \kappa.
$$
It remains to show that for any  action $i \in \{1,2\}$, the agent's utility under $\langle \tau, 3 \rangle$ is at most $\kappa$.
If $t(r_i) < t'(r_i)$, then $c_i = t(r_i) - \kappa$, and the agent's expected utility for action $i$ is $U_A(i \mid \tau) = U_A(i \mid t) = t(r_i) - c_i = \kappa.$
Similarly, if $t'(r_i) \leq t(r_i)$, then $c_i = t'(r_i) - \kappa$, and the agent's utility for action $i$ is $U_A(i \mid \tau) = U_A(i \mid t') = t'(r_i)-c_i = \kappa$.
So in either case, the agent's utility from the ambiguous contract is at most $\kappa$.

We complete the argument by showing that action $3$ cannot be implemented by any classic 
contract (even if we don't restrict the classic 
contract to come from class $\classOfContracts$). 
We have 
$$
    \sum_{j=1}^2 q_j c_j = \left(\sum_{j=1}^2{q_j \cdot \min(t(r_j),t'(r_j))}\right) - \kappa < \left(\sum_{j=1}^2{q_j \cdot t(r_j)}\right) - \kappa = c_{3}.
$$
The convex combination of actions $1, 2$ via vector $(q_1, q_2)$ thus yields the same distribution over rewards as action $3$, but at a strictly lower cost. 
By Proposition~\ref{prop:implementable}, this means that action $3$ is not implementable by any classic 
contract.
\end{proof}

As an immediate corollary of our characterization we obtain that
for any fixed $d \in \mathbb{R}_+$ and $\beta\le r_1$, the class of payment functions  $\classOfContracts_{d}(\beta) = \{t(x) = \alpha \cdot (x-\beta)^d \mid \alpha \ge 0\}$ is ambiguity-proof.
If all rewards are non-negative, then we can set $\beta=0$ to see that the class of linear contracts $\classOfContracts_{1}(0)$ is ambiguity-proof.  For general rewards, if $\beta<0$, then $\classOfContracts_{1}(\beta)$ describes an ambiguity-proof class of affine contracts.  
This is cast in the following corollary.

\begin{corollary}\label{cor:ambiguity-proof}
For any fixed $d \in \mathbb{R}_+$, and $\beta\le r_1$, the class of payment functions  $\classOfContracts_{d}(\beta) = \{t(x) = \alpha \cdot (x-\beta)^d \mid \alpha \ge 0\}$ is ambiguity-proof.  
In particular, when all rewards are positive, the class of linear payment functions  (corresponding to $\classOfContracts_1(0)$) is ambiguity-proof.
\end{corollary}

On the other hand, our characterization implies that many natural classes of contracts, such as the class of all affine contracts, all polynomial contracts, or the class of all monotone contracts, fail to be ambiguity-proof.

\section{Mixing Hedges Against Ambiguity}\label{sec:mixing}

In this section we explore the power of ambiguity when the agent is allowed to select a mixed action and the principal is allowed to implementing mixed actions. Our main result (Theorem~\ref{thm:mixing-hegdes-against-ambiguity}) is that in this case, the principal cannot gain from using an ambiguous contract. 
\cite{Bade23} obtains a similarly-spirited result in a mechanism design context, showing that if agents are dynamically consistent, meaning that they update their beliefs in response to information so as to make it optimal to continue with their ex-ante optimal plan of action, then ambiguity does not expand the set of implementable social choice functions.  
In contrast, \cite{Kambhampati23} shows that a principal who entertains ambiguous beliefs about the actions available to an agent can typically improve her payoff by offering a random contract to the agent.

\vspace*{-8pt}

\subsection{Extension  to Mixed Actions} 

The definition of a payment function remains the same, $t = (t_1, \ldots, t_m) \in \mathbb{R}^m_{+}$ defines a non-negative transfer $t_j$ for each outcome $j \in [m]$.  A mixed action of the agent, denoted by $\psi \in \Delta^n$, is a convex combination over actions $i \in [n]$, so that $\psi_i$ denotes the probability with which the agent chooses action $i$. A pure action is the special case of a  mixed action in which $\psi_i = 1$ for some $i$ and $\psi_{i'} = 0$ for all other $i' \neq i$. 
The expected reward of mixed action $\psi$, is $R_\psi = \sum_{i=1}^{n} \psi_i  R_i$. 
We write $T_{\psi}(t) = \sum_{i=1}^{n} \psi_i {T}_i(t)$ for the expected payment for mixed action $\psi$ under payment function $t$.

The agent's expected utility for mixed action $\psi'$ under classic contract $\langle t, \psi \rangle$ is 
$\UAmid{\psi'}{t} = \sum_{i=1}^{n} \psi'_i  \UAmid{i}{t}$.
Classic contract $\langle t, \psi\rangle$ is incentive compatible if for any mixed action $\psi'$, it holds that $\UAmid{\psi'}{t}\leq \UAmid{\psi}{t}.$ We use $\UA{\langle t, \psi\rangle}$ for the agent's utility under incentive compatible contract $\langle t, \psi\rangle$, and $\UP{\langle t, \psi\rangle} = \sum_i \psi_i \UPmid{i}{t}$ for the principal's utility.

We define the agent's expected utility for mixed action $\psi'$ under ambiguous contract $\langle \tau,\psi \rangle$ to be the minimum utility under any payment function $t \in \tau$. That is, the agent's expected utility is $\UAmid{\psi'}{\tau} = \min_{t \in \tau} \UAmid{ \psi'}{t}$.

We say that ambiguous contract $\langle \tau,\psi \rangle$ implements mixed action $\psi$ if for every mixed action $\psi'$ it holds that $\UAmid{\psi'}{\tau} \leq \UAmid{\psi}{\tau}$.
An incentive compatible contract  
$\langle \tau,\psi \rangle$ is consistent if for any two contracts $t,t' \in \tau$ it holds that
$\UPmid{\psi}{t} = \UPmid{\psi}{t'}$ 
and, thus, $T_{\psi}(t) = T_{\psi}(t')$.
For a (consistent) incentive compatible contract $\langle \tau, \psi\rangle$, we write $\UA{\langle \tau, \psi\rangle}$ for the agent's utility under $\psi$, and  $T_{\psi}(\tau)$ for the resulting expected payment.
The principal's expected utility under (consistent) incentive compatible contract $\langle \tau, \psi \rangle$ is 
$\UP{\langle \tau , \psi \rangle} = \UPmid{\psi}{\tau} = R_{\psi} - T_{\psi}(\tau)$.

In what follows, without loss, we restrict attention to consistent incentive compatible 
contracts.

\subsection{Mixing Hedges Against Ambiguity}

When the agent can choose mixed actions, the principal cannot gain by employing an ambiguous contract. To prove this result we make use of the min-max theorem applied to a suitably defined zero-sum game.

\begin{theorem}\label{thm:mixing-hegdes-against-ambiguity}
Consider an incentive compatible ambiguous contract $\langle \tau,\psi \rangle$, with payoffs $\UAmid{\psi}{\tau}$ and $\UPmid{\psi}{\tau}$.
Then there exists an incentive compatible classic contract $\langle \hat t, \psi \rangle$ with the same agent and principal payoffs. 
\end{theorem}

\begin{proof}
Let $\langle \tau,\psi \rangle$ be an incentive compatible ambiguous contract with payoffs $\UAmid{\psi}{\tau}$ and $\UPmid{\psi}{\tau}$.

Consider a zero-sum game played by the agent and the principal. The agent's (convex and compact) strategy set is the set of mixed actions $\psi'\in \Delta^n$, while the principal's (also convex and compact) strategy set is the set of payment functions $\hat{\mathcal T}=\{t\in \mathbb{R}_+^m:\sum_{i=1}^n\sum_{j=1}^m\psi_i \prob{i}{j} t_j=T_{\psi}(\tau)\}$. 
This is the set of payment functions 
that preserve the principal's payoff $\UPmid{\psi}{\tau}$ when the agent plays action $\psi$.
Note that $\hat{\mathcal T}$ is non-empty since every payment function $t \in \tau$ has the same expected payment under $\psi$ (namely $T_{\psi}(\tau)$).
The agent's payoff in this game is  $\sum_{i=1}^n\sum_{j=1}^m\psi_i(p_{ij}t_j-c_i)$, while the principal's payoff is the negative of this quantity. Note that this way the agent's payoff in this game is precisely the agent's utility in the principal-agent setting, while the principal's payoff differs from that in the principal-agent setting.

The min-max theorem implies:
\begin{equation}\label{winken}
\max_{\psi'\in\Delta^n}\min_{t\in \hat{\mathcal T}}\sum_{i=1}^n\sum_{j=1}^m\psi'_i(p_{ij}t_j-c_i) = \min_{t\in \hat{\mathcal T}}\max_{\psi'\in\Delta^n}\sum_{i=1}^n\sum_{j=1}^m\psi'_i(p_{ij}t_j-c_i).
\end{equation}

Note that $\hat{\mathcal{T}}$ is just a set of payment functions, so we can interpret it as a (potentially infinite-size) set that constitutes the payment functions within an ambiguous contract. 
Moreover, by the construction of $\hat {\mathcal T}$, we have $\UAmid{\psi}{\hat {\mathcal T}}  = \UAmid{\psi}{\tau}$, since 
$\psi$ ensures utility 
$\UPmid{\psi}{\tau}$ against every element of $\hat {\mathcal T}$ to the principal,
thus utility $\UAmid{\psi}{\tau}$ against every element of $\hat {\mathcal T}$ to the agent.

We next show that the value of the zero-sum game is also equal to this quantity. 
To this end, we show that 
\[
\UAmid{\psi}{\tau} \geq 
\max_{\psi'\in\Delta^n}\min_{t\in \hat{\mathcal T}}\sum_{i=1}^n\sum_{j=1}^m\psi'_i(\prob{i}{j} t_j-c_i) \geq 
\UAmid{\psi}{\hat {\mathcal T}}.
\]
To see this, observe that
\begin{eqnarray}
\UAmid{\psi}{\tau} &=& \min_{t\in \tau}\sum_{i=1}^n\sum_{j=1}^m\psi_i(\prob{i}{j} t_j-c_i)~=~\max_{\psi'\in\Delta^n}\min_{t\in \tau}\sum_{i=1}^n\sum_{j=1}^m\psi'_i(\prob{i}{j}t_j-c_i)\label{blinken}\\
&\ge&\max_{\psi'\in\Delta^n}\min_{t\in \hat{\mathcal T}}\sum_{i=1}^n\sum_{j=1}^m\psi'_i(\prob{i}{j}t_j-c_i)\label{nodd} 
~\geq~ \min_{t\in \hat{\mathcal T}}\sum_{i=1}^n\sum_{j=1}^m\psi_i(\prob{i}{j}t_j-c_i)\\
& = & \UAmid{\psi}{\hat {\mathcal T}}.\label{last}
\end{eqnarray}
The second equality in \eqref{blinken} holds by the fact that $\tau$ implements $\psi$, the first inequality in~\eqref{nodd} follows from comparing feasible sets for the corresponding minimizations, and the following inequality holds because $\psi$ is feasible in the maximization.

Consider the payment function $\hat t \in \argmin_{t\in \hat{\mathcal T}}\max_{\psi'\in\Delta^n}\sum_{i=1}^n\sum_{j=1}^m\psi'_i(\prob{i}{j}t_j-c_i)$.  From \eqref{winken}, there is no action in $\Delta^n$ giving the agent a payoff against $\hat t$ higher than 
$\max_{\psi'\in\Delta^n}\min_{t\in \hat{\mathcal T}}\sum_{i=1}^n\sum_{j=1}^m\psi'_i(p_{ij}t_j-c_i)=
\UAmid{\psi}{\tau}$.  By construction, 
mixed action $\psi$ gives this payoff when facing the payment function $\hat t$.  Hence,  $\langle \hat t, \psi \rangle$ is an incentive compatible classic contract with payoffs 
$\UAmid{\psi}{\tau}$ to the agent and $\UPmid{\psi}{\tau}$ to the principal.
\end{proof}

The ability to mix provides the agent with more alternative actions, tightening the incentive constraints enough to dissipate any advantage the principal gains from ambiguous contracts.  The following example illustrates this.

\begin{example}Return to Example \ref{ex:improve}. 
{\em
Suppose the agent is restricted to pure actions.  As we have seen, the uniquely optimal ambiguous contract is 
$\langle \tau,4 \rangle$, with 
$\tau = \{(0,3/2,0),(0,0,3/2)\}$.
The payoffs to actions $1$, $2$, $3$ and $4$ under $\tau$ are $0$, $-1/4$, $-1/4$, and $0$.  Pure actions $2$ and $3$ are thus strictly inferior to action $4$ for the agent. 

Now suppose the agent can choose a mixed action.  If the payoffs to the pure actions had been generated by classic contracts, then no mixture over actions $2$ and $3$ could give a payoff higher than $-1/4$, and hence no such mixture could be superior to action 4.  Under ambiguity, this familiar property of mixed actions breaks down.  The mixture that places probability $1/2$ on each of actions $2$ and $3$ gives an expected payoff of $1/8$, strictly larger than the payoffs to any pure strategy in the support of the mixture, ensuring that the ambiguous contract
$\langle \tau, 4 \rangle$ 
no longer implements action $4$.  Indeed, given the agent's ability to choose this mixture, action $4$ cannot be implemented at any payment less than $1$, matching the payment under classic contracts.
\hfill\rule{.1in}{.1in}}
\end{example}

\subsection{Relation to Ellsberg Paradox and Raiffa's Critique}

\cite{Ellsberg61} pioneered the argument that humans tend to prefer choices with quantifiable risks over those with unquantifiable, incalculable risks, giving rise to  the ambiguity-aversion literature.  We have shown that the principal can take advantage of the agent's ambiguity-aversion if, but only if, the agent is restricted to pure actions. 

A critique of Ellsberg's experiments, raised by \cite{Raiffa61}, is that when faced with the Ellsberg urns, a player could mentally flip a coin  and implement a mixed action that induces an objective probability distribution over outcomes.  Doing so removes all of the ambiguity from the decision, and with it any need for ambiguity aversion.  
Raiffa's argument highlights the potential power an ambiguity-averse agent can derive from engaging in mixed actions. Indeed, Theorem~\ref{thm:mixing-hegdes-against-ambiguity} shows 
that engaging in mixed strategies completely eliminates the principal's power stemming from ambiguous contracts.

Raiffa's argument has given rise to a discussion, centered around the question of whether such mental coin flips are indeed effective in banishing uncertainty (e.g., \cite{kezhang2020,saito2015}).
One way of formalizing this ``critique of Raiffa's critique,''  is \cite{Bade23}'s notion of  dynamic semi-consistency. According to this behavioral assumption, agents should not update their beliefs in response to signals that are independent of the environment (such as an independent coin flips). This notion rules out the type of mixing required for Theorem~\ref{thm:mixing-hegdes-against-ambiguity}, but aligns with our behavioral assumptions underlying the results in Sections~\ref{sec:model}--\ref{sec:Ambiguity Proofness}.

\bibliographystyle{apalike}
\bibliography{bibliography}
\newpage
\appendix

\input{appendix}

\end{document}

%% file: macros.tex
\newcommand{\instance}{$(c,r,p)$\xspace}
\newcommand{\contract}{\langle t,i \rangle}
\newcommand{\ambcontract}{\langle \tau, i \rangle}
\newcommand{\ambcontractfull}{\langle \{t^1, \ldots, t^k\},i \rangle}
\newcommand{\ambcontractprime}{\langle \tau', i \rangle}
\newcommand{\UA}[1]{U_A(#1)}
\newcommand{\UAmid}[2]{U_A({#1} \mid {#2})}
\newcommand{\UP}[1]{U_P(#1)}
\newcommand{\UPmid}[2]{U_P({#1} \mid {#2})}
\newcommand{\Reward}{R}
\newcommand{\Welfare}{W}
\newcommand{\Payment}[2]{T_{#1}(#2)}

\newcommand{\prob}[2]{p_{{#1}{#2}}}

\newcommand{\payment}[2]{T_{#1}({#2})}

\newcommand{\principalAgentSetting}{$(c,r,p)$\xspace}

\newcommand{\jHat}{\hat{j}}
\newcommand{\tHat}{\hat{t}}

\newcommand{\classOfContracts}{\mathcal{T}\xspace}

%% file: appendix.tex
\section{Proof of Proposition~\ref{prop:implementable}}\label{app:implementable}

\begin{figure}[t]
\begin{minipage}[t]{0.55\textwidth}
$
\begin{array}{l}
\min\hspace*{1pt}\displaystyle{\sum_jp_{ij}t_j}\\  
{\rm s.t.~~}\displaystyle{\sum_jp_{ij}t_j-c_i\ge\sum_jp_{i'j}t_j-c_{i'}~~~~\forall i'\neq i}\\
~~~~~~\hspace*{1pt}\displaystyle{t_j\ge 0 \qquad\qquad\qquad\qquad\qquad\quad~\hspace*{2pt}\forall j}\\
\end{array}
$
\end{minipage}
\vrule~
\begin{minipage}{0.5\textwidth}
$
\begin{array}{l}
\max\displaystyle{\sum_{i'\neq i}\lambda_{i'}(c_i-c_{i'})}\\  
{\rm s.t.~~}\displaystyle{\sum_{i'\neq i}\lambda_{i'}(p_{ij}-p_{i'j})\le p_{ij}~~~~\forall j}\\
~~~~~~\displaystyle{\lambda_{i'}\ge 0 \qquad\qquad\qquad~~~~~~\forall i'\neq i}\\
\end{array}
$
\end{minipage}

\vspace{10pt}
\begin{minipage}{\textwidth}
\hspace*{40pt}(a) Primal LP for action $i$. \hspace*{90pt} (b) Dual LP for action $i$.
\end{minipage}
\caption{The minimum payment LP.}
\label{fig:minpaylp}
\end{figure}

\begin{proof}[Proof of Proposition~\ref{prop:implementable}]
Figure \ref{fig:minpaylp} shows the linear program determining the optimal classic contract implementing action $i$, and its dual.  Consider the primal LP in Figure~\ref{fig:minpaylp} (a) for action $i$, but with the objective $\min \sum_j p_{ij} t_j$ replaced with  $\min 0$. Action $i$ is implementable if and only if the resulting LP is feasible. The dual of this LP is:
\begin{align*}
\max \quad & \sum_{i' \neq i} \lambda_{i'} (c_i-c_{i'}) \\
& \sum_{i' \neq i} \lambda_{i'} (p_{ij} - p_{i'j}) \leq 0 && \forall~j\\
&\lambda_{i'} \geq 0 &&\forall~i' \neq i
\end{align*}

By strong duality for a general primal-dual pair we can have the following four cases: 

(1)~The dual LP and the primal LP are both feasible. 

(2)~The dual LP is unbounded and the primal LP is infeasible. 

(3) The dual LP is infeasible and the primal LP is unbounded. 

(4) The dual LP and the primal LP are both infeasible. 

\noindent In our case the dual LP is always feasible (we can choose $\lambda_{i'} = 0$ for all $i' \neq i$). This rules out cases (3) and (4). So in order to prove the claim it suffices to show that the dual LP is unbounded if and only if there exists a convex combination $(\gamma_{i'})_{i'\neq i}$
such that $\sum_{i' \neq i} \gamma_{i'} p_{i'j} = p_{ij}$ for all $j$ and $\sum_{i' \neq i} \gamma_{i'} c_{i'} < c_i$.

We first show that if such a convex combination exists, then the dual LP is unbounded. Indeed, if such a convex combination exists, then it corresponds to a feasible solution to the dual LP because, for all $j$, 
\[
\sum_{i' \neq i} \gamma_{i'} (p_{ij} - p_{i'j}) = \left(\sum_{i' \neq i} \gamma_{i'}\right) p_{ij} - \left(\sum_{i' \neq i} \gamma_{i'} p_{i'j}\right) 
= 0,
\]
where we used that $\sum_{i'\neq i} \gamma_{i'}=1$ and $\sum_{i'\neq i} \gamma _{i'}p_{i'j} = p_{ij}.$
Moreover, the objective value achieved by this solution is 
\[
\sum_{i' \neq i} \gamma_{i'} (c_i-c_{i'}) = \left(\sum_{i' \neq i} \gamma_{i'}\right) c_i - \left(\sum_{i' \neq i} \gamma_{i'} c_{i'}\right) 
= \delta
\]
for some $\delta > 0$, where we used that $\sum_{i' \neq i}\gamma_{i'}=1$ and $\sum_{i'\neq i} \gamma_{i'} c_{i'} < c_i.$
But then for any $\kappa \geq 0$ setting the dual variables to $\kappa \cdot \gamma_{i'}$ for $i' \neq i$ results in a feasible solution whose objective value is equal to $\kappa \cdot \delta$. So the dual LP is unbounded.

We next show that if the dual LP is unbounded, then a convex combination with the desired properties must exist. Since the dual LP is unbounded, for any $\delta > 0$ there must be a feasible solution to the dual LP,  $(\lambda_{i'})_{i'\neq i}$, such that $\sum_{i' \neq i} \lambda_{i'} (c_i - c_{i'}) \geq \delta$ and $\sum_{i' \neq i} \lambda_{i'} (p_{ij} - p_{i'j}) \leq 0$ for all $j$. Now consider $\gamma_{i'} = \lambda_{i'}/(\sum_{i' \neq i} \lambda_{i'})$ for all $i' \neq i$. We claim that $(\gamma_{i'})_{i'\neq i}$ is 
a convex combination with the desired properties. First note that $(\gamma_{i'})_{i'\neq i}$ is 
a convex combination, i.e., $\gamma_{i'} \in [0,1]$ for all $i' \neq i$ and $\sum_{i' \neq i} \gamma_{i'} = 1$. Also note that
\[
\sum_{i' \neq i} \gamma_{i'} (c_i - c_{i'}) = \frac{1}{\sum_{i' \neq i}\lambda_{i'}} \sum_{i' \neq i} \lambda_{i'} (c_i - c_{i'}) = \frac{1}{\sum_{i' \neq i}\lambda_{i'}} \cdot \delta > 0
\]
and therefore $\sum_{i' \neq i} \gamma_{i'} c_{i'} < (\sum_{i' \neq i} \gamma_{i'}) c_i = c_i$. Moreover, for all $j$, using the fact that $\sum_{i' \neq i} \lambda_{i'} p_{i'j} \geq (\sum_{i' \neq i} \lambda_{i'}) p_{ij}$, we must have
\[
\sum_{i' \neq i} \gamma_{i'} p_{i' j} = \frac{1}{\sum_{i' \neq i} \lambda_{i'}}\sum_{i' \neq i} \lambda_{i'} p_{i' j} \geq \frac{1}{\sum_{i' \neq i} \lambda_{i'}} \left(\sum_{i' \neq i} \lambda_{i'}\right) p_{i j} = p_{i j}.
\]
So we know that for all $j$, $\sum_{i' \neq i} \gamma_{i'} p_{i' j} \geq p_{i j}$. We claim that, for all $j$, this inequality must hold with equality. Indeed, assume for contradiction that for some $j'$ we have a strict inequality. By summing over all $j$, we then have
\begin{align}
\sum_{j} \left(\sum_{i' \neq i} \gamma_{i'} p_{i' j}\right) &> \sum_j p_{i j} = 1,
\label{eq:more-than-one}
\end{align}
where we used that $p_i$ is a probability distribution over outcomes $j$. 
On the other hand, we have that
\begin{align}
\sum_{j} \left(\sum_{i' \neq i} \gamma_{i'} p_{i' j}\right) = \sum_{i' \neq i} \gamma_{i'} \left(\sum_{j}  p_{i' j}\right) = \sum_{i' \neq i} \gamma_{i'} = 1,
\label{eq:exactly-one}
\end{align}
where we used that the $p_{i'}$'s are also probability distributions over outcomes $j$ and that $(\gamma_{i'})_{i'\neq i}$ is a convex combination. Combining \eqref{eq:more-than-one} with \eqref{eq:exactly-one} we get the desired contradiction.
\end{proof}

\section{Tightness of Theorem~\ref{thm:wlogSOP}}\label{app:sop-tight}

\begin{proposition}\label{prop:sop-tight}
Assume $m > 4$. There exist instances where the best ambiguous contract composed of $n-1 = m$ payment functions is strictly better than any ambiguous contract that is composed of fewer payment functions.
\end{proposition}

\begin{figure}
\begin{table}[H]
    \centering
    \scalebox{0.9}{
    \begin{tabular}{|c|c|c|c|c|c|c|c|c|c|}
    \hline
        \rule{0pt}{3ex} \hspace{2.0mm} rewards:  \hspace{2.0mm}&\hspace{2.0mm} $r_1 = 0$ \hspace{2.0mm}&\hspace{2.0mm} $r_2 = \delta$ \hspace{2.0mm}&\hspace{2.0mm} $r_3 = 2\delta$ \hspace{2.0mm}&\hspace{2.0mm} $\dots$ \hspace{2.0mm}&\hspace{2.0mm} $r_m = m-1$ \hspace{2.0mm}&\hspace{2.0mm} costs \hspace{2.0mm} \\[1ex] \hline
        \rule{0pt}{3ex}  action $1$: & $ 0$ & $\frac{1}{m-1}$ & $\frac{1}{m-1}$ & $\dots$ & $\frac{1}{m-1}$ & $c_1 =0$  \\[1ex] 
         \rule{0pt}{3ex}  action $2$:  & $ \frac{1}{m-1}$ & $0$ & $\frac{1}{m-1}$ & $\dots$ & $\frac{1}{m-1}$ & $c_2 = 0$  \\[1ex] 
         \rule{0pt}{3ex}  action $3$:  & $ \frac{1}{m-1}$ & $\frac{1}{m-1}$  & $ 0$ & $\dots$ & $\frac{1}{m-1}$  & $c_3 = 0$  \\[1ex] 
         \rule{0pt}{3ex}  $\vdots$ & $\vdots$ & $\vdots$ & $\vdots$ & $\ddots$ &  $\vdots$ & $\vdots$  \\[1ex] 
         \rule{0pt}{3ex}  action $m$:  & $ \frac{1}{m-1}$  & $ \frac{1}{m-1}$  & $\frac{1}{m-1}$  &  $\dots$ & $0$ & $c_m = 0$  \\[1ex] 
         \rule{0pt}{3ex}  action $m+1$:  & $\frac{1}{(m-1)^2}$ & $\frac{1}{(m-1)^2}$ &  $\frac{1}{(m-1)^2}$ &  $\dots$ & $1-\frac{1}{m-1}$ & $c_{m+1} = 1$  \\[1ex] \hline
    \end{tabular}
    }
\end{table}
\caption{Instance $(c,r,p)$ used in the proof of Proposition~\ref{prop:sop-tight}.}
\label{fig:sop-tight}
\end{figure}

\begin{proof}
Consider the instance $(c,r,p)$ depicted in Figure~\ref{fig:sop-tight} with $m$ outcomes and $n = m+1$ actions, where $\delta$ is a vanishingly small positive number. 
First note that in this instance the expected reward $R_i$ of any action $i \leq m$ is 
$$ 
    R_i 
    \leq \sum_{j =1}^{m} \frac{1}{m-1} r_j = \left(\sum_{j =1}^{m-1} \frac{(j-1)}{m-1} \cdot \delta \right) + \frac{m-1}{m-1}= 1 + \frac{1}{2} (m^2-3m+2) \delta,
$$
which can be made arbitrarily close to $1$ by choosing $\delta$ small enough.
On the other hand, the expected reward of action $m+1$ is
\[
R_{m+1} = \left( \sum_{j =1}^{m-1} \frac{1}{(m-1)^2} \cdot (j-1)\delta \right) + \left(1-\frac{1}{m-1}\right) \cdot (m-1) = (m-2) \left(1+\frac{1}{2(m-1)}\delta\right).
\]
Since $\delta$ is positive and $m > 4$, it follows that $R_{m+1} \geq m-2$ and $W_{m+1} = R_{m+1} - c_{m+1} \geq m-3 \geq 2$. 
To prove the claim it thus suffices to show that there exists a (consistent) IC ambiguous contract consisting of $n-1 = m$ payment functions that implements action $m+1$ with an expected payment equal to $c_{m+1}$; while any (consistent) IC ambiguous contract that consists of strictly fewer payment functions must pay strictly more in order to implement action $m+1$. 

We first show that we can indeed implement action $m+1$ with an ambiguous contract consisting of $n-1 = m$ payment functions with an expected payment equal to $c_{m+1}$. To this end consider the ambiguous contract $\langle \tau, m+1\rangle$ consisting of $n-1 = m$ SOP payment functions:
\begin{eqnarray*}
    \tau & = & \{t^1,t^2,\dots,t^m\} \hspace{3.0mm} s.t\\
    t^j_j & = & \frac{c_{m+1}}{\prob{(m+1)}{j}} \hspace{3.0mm} \text{and} \hspace{3.0mm} t^j_{j'} = 0 \hspace{3.0mm} \forall j' \ne j
\end{eqnarray*}
It is then easy to verify, that for any payment function $t^j \in \tau$, $T_{m+1}(t^j) = c_{m+1}$, which shows that $\langle \tau, m+1 \rangle$ is consistent. Since $T_{m+1}(\tau) = c_{m+1}$, the agent's utility for action $m+1$ is $\UAmid{m+1}{\tau} = 0$. On the other hand, for any action $i \neq m+1$, there is an SOP contract  $t^j \in \tau$ such that 
$\payment{i}{t^j} = 0$. Thus, for any action $i \neq m+1$, $\UAmid{i}{\tau} \leq 0$, which shows that $\langle \tau, m+1 \rangle$ is IC. 

To complete the proof, assume by contradiction that there is a (consistent and) IC ambiguous contract $\langle\tau',m+1\rangle$ with $T_{m+1}(\tau') = c_{m+1}$ but $|\tau'| < n-1 = m$. 
Since $\UAmid{m+1}{\tau'} = 0$, in order for $\langle \tau', m+1 \rangle$ to implement action $m+1$, we must have $\UAmid{i}{\tau'} \leq 0$ for each action $i \in [m]$. The only way to achieve this is to have, for each such action $i \in [m]$, a payment function $t' \in \tau'$ under which $T_{i'}(t') = 0$. Since for each action $i \in [m]$, there exist only one outcome, denoted by $j(i)$, for which it has 0 probability, and for all other $j'$ we have that $\prob{i}{j'} > 0$, we get that the only type of payment function for which its expected payment is $0$, is an SOP payment function that pays for that specific outcome, $j(i)$. Since each action has a unique outcome for which it has $0$ probability (i.e., for all $i,i' \in [m]$, $i \neq i'$, it holds that $j(i)\neq j(i')$), we get that for each action the SOP payment function for which its expected payment is $0$, is unique as well. This shows that $\tau'$ must consist of at least $n-1 = m$ payment functions, in contradiction to our assumption.
\end{proof}

\section{Proof of Proposition~\ref{prop:opt-ambiguous}}\label{app:opt-ambiguous}

\begin{proof}[Proof of Proposition~\ref{prop:opt-ambiguous}]
By Proposition~\ref{hadyn} action $i$ is implementable if and only if there is no other action $i'$ such that $p_{i'} = p_i$ and $c_{i'} < c_i$. Therefore, if $A = \emptyset$, then it must hold that $c_i = 0$. In this case, $\langle \{(0,\ldots,0)\},i \rangle$ implements action $i$, and clearly no other contract can do so with a lower expected payment. 
So suppose $A \neq \emptyset$, and consider the contract $\langle \tau, i\rangle$ for this case from the statement of the proposition. We have already argued that this contract is (consistent and) IC (in the proof of Proposition~\ref{hadyn}). It remains to show that it is optimal.

For a contradiction suppose that there exists a (consistent) IC contract $\langle \tau',i \rangle$, 
such that $\Payment{i}{\tau'} < \Payment{i}{\tau}$. 
By Theorem~\ref{thm:wlogSOP}, we can assume that $\langle \tau' , i \rangle$ is an SOP contract. 
 
Consider an action $i' \in A$ for which
\begin{align}
\min \left\{x \geq 0 \;\middle|\; p_{i j(i')} \cdot \frac{x}{p_{ij(i')}} - c_i \geq p_{i'j(i')} \cdot \frac{x}{p_{ij(i')}} - c_{i'} \right\} = T,
\label{eq:T}
\end{align}
and recall that $T_i(\tau) = T$.

First note that since $\langle \tau,i \rangle$ is IC, we must have $T \geq c_i$. Moreover, we cannot have $T = c_i$, because then we would have $\Payment{i}{\tau'} < \Payment{i}{\tau} = T = c_i$, which would contradict our assumption that $\langle \tau',i \rangle$ is IC. 

So consider the case where $T > c_i$. In this case, since $p_{ij(i')}/p_{ij(i')} > p_{i'j(i')}/p_{ij(i')}$, Equation~\eqref{eq:T} implies that $c_i > c_{i'}$.

Since $\langle \tau', i \rangle$ is IC, there must be an SOP payment function $t' \in \tau'$ such that $\UAmid{i'}{\tau'} = \UAmid{i'}{t'} \leq \UAmid{i}{t'} = \UAmid{i}{\tau'}$. Since $t'$ is SOP, there must be an outcome $j'$ for which $t'_{j'} \geq 0$, while $t'_{j''} = 0$ for all $j'' \neq j'$.

If $j' = j(i')$, then 
\[
p_{i'j(i')} \cdot \frac{T_i(\tau')}{p_{ij(i')}} - c_{i'}
= 
U_A(i' \mid t') 
\leq 
U_A(i \mid t') 
=
p_{ij(i')} \cdot \frac{T_i(\tau')}{p_{ij(i')}} - c_i,
\]
which contradicts the minimality of $T$. 


Otherwise, $j' \neq j(i')$. 
In this case, since $U_A(i \mid t') \geq U_A(i' \mid t')$ and $c_i > c_{i'}$, we must have $p_{i'j'}/p_{ij'}< 1$. Moreover, by definition of $j(i')$, we must have
$p_{i'j'}/p_{ij'} \geq p_{i'j(i')}/p_{ij(i')}$. 
So
\[
T_i(\tau) = \frac{c_i - c_{i'}}{1-\frac{p_{i'j(i')}}{p_{ij(i')}}} \leq \frac{c_i - c_{i'}}{1-\frac{p_{i'j'}}{p_{ij'}}} \leq T_i(\tau'),
\]
where 
the equality holds by definition of $T$, the first inequality uses that $c_i - c_{i'} > 0$, and the final inequality holds because  $U_A(i \mid t') \geq U_A(i' \mid t')$. We obtain a contradiction to our assumption that $T_i(\tau') < T_i(\tau)$.
\end{proof}

\section{Proof of Theorem~\ref{theo:unified MLRP SOP}}\label{app:proof-of-unified-mlrp-sop}

\begin{proof}
Theorem \ref{thm:wlogSOP} ensures that there exists an ambiguous contract $\langle \tau'', i \rangle$ consisting of the SOP payment functions $\tau'' = \{t^1,\ldots, t^k\}$ that implements action $i$ with expected payment $T_{i}(\tau)$, and with $t^1$ and $t^k$ specified as in point 2.  To prove the result, it suffices to show that the ambiguous contract $\langle \tau', i \rangle$ with $\tau' = \{t^1,t^k\}$ also implements action $i$.  In turn, it suffices for this result to note that
\begin{eqnarray}
i'<i&\implies &T_{i'}(\tau'')=\min_{j=1,\ldots, k} \prob{i'}{j} \frac{T_{i}(\tau)}{\prob{i}{ j}}\ge \prob{i'}{h} \frac{T_{i}(\tau)}{\prob{i}{h}}\ge \min_{j=1, k} \prob{i'}{j} \frac{T_{i}(\tau)}{\prob{i}{ j}}=T_{i'}(\tau'),~~~~\label{uriah}\\
i'>i&\implies &T_{i'}(\tau'')=\min_{j=1,\ldots k}p_{i'j} \frac{T_{i}(\tau)}{\prob{i}{ j}}\ge \prob{i'}{\ell} \frac{T_{i}(\tau)}{\prob{i}{\ell}}\ge\min_{j=1, k}\prob{i'}{j}\frac{T_{i}(\tau)}{\prob{i}{j}}=T_{i'}(\tau').~~~~\label{heap}
\end{eqnarray}
The first inequality in each of these statements follows from the MLRP condition.  In the case of \eqref{uriah}, for example, this inequality follows from noting that if $i'<i$, then the MLRP condition implies that (since $h>j$) 
$p_{ih}p_{i'j}\ge p_{ij}p_{i'h}$. 
\end{proof}

\section{Proof of Theorem~\ref{theo:unified MLRP monotone ambiguous contract}}\label{app:montone-with-mrlp}

\begin{proof}[Proof of Theorem~\ref{theo:unified MLRP monotone ambiguous contract}]
By construction, the ambiguous contract $\langle \tau', i \rangle$ with $\tau' = \{t^1,t^k\}$ is consistent and $\payment{i}{\tau'} = \payment{i}{\tau}$, so point 1 is satisfied. To show that $\langle \tau',i\rangle$ is incentive compatible, we note that since $\langle \tau, i \rangle$ implements action $i$, 
each action $i' \neq i$ has a monotone payment function $t^{(i')} \in \tau$ such that $\UAmid{i'}{t^{(i')}} \le \UAmid{i}{\tau}$.
It then suffices to show that each action $i' \neq i$ has a payment function $t \in \tau'$ for which 
$\payment{i'}{t} \leq \payment{i'}{t^{(i')}}$.

For actions $i' \neq i$ such that $c_{i'} \le c_{i}$, we claim that $\payment{i'}{t^k} \leq \payment{i'}{t^{(i')}}$. Indeed, for any such action $i'$,
\begin{align*}
\payment{i'}{t^k} = \sum_{j = 1}^{m} \prob{i'}{j} \cdot t^k_j = \sum_{j = h}^{m} \prob{i'}{j} \cdot t^k_j = \prob{i'}{h} \cdot t_h^k,
\end{align*}
where we used that $t^k_j = 0$ for $j < h$, and that $\prob{i'}{j} = 0$ for $j > h$ by the MLRP condition. Substituting the definition of $t_h^k = \payment{i}{\tau}/\prob{i}{h}$ and using that $\payment{i}{\tau} = \payment{i}{t^{(i')}}$ by consistency, we obtain 
\begin{align*}
    \payment{i'}{t^k} 
    = \prob{i'}{h} \cdot \frac{\payment{i}{t^{(i')}}}{\prob{i}{h}} =  \prob{i'}{h} \cdot \frac{\sum_{j \in [m]}{\prob{i}{j}\cdot t^{(i')}_j}}{\prob{i}{h}} 
    =  \sum_{j = 1}^h{ \frac{\prob{i'}{h} \cdot \prob{i}{j}}{\prob{i}{h}}\cdot t^{(i')}_j },
\end{align*}
where for the last step we used that $\prob{i}{j} = 0$ for $j >h$ (by definition of $h$).

By the MLRP condition, since $c_{i'} \le c_{i}$, for all $j \le h$: 
\begin{align*}
    \frac{\prob{i'}{h}}{\prob{i}{h}} \le \frac{\prob{i'}{j}}{\prob{i}{j}} \Longrightarrow \frac{\prob{i'}{h} \cdot \prob{i}{j}}{\prob{i}{h}} \le \prob{i'}{j}.
\end{align*}
Using this we obtain
\begin{align*}
\payment{i'}{t^k} 
    \leq \sum_{j=1}^{h} p_{i'j} \cdot t_j^{(i')} = \sum_{j=1}^{m} p_{i'j} \cdot t_j^{(i')} = \payment{i'}{t^{(i')}},
\end{align*}
where we again used that $\prob{i'}{j} = 0$ for $j > h$ by the MLRP condition.

For actions $i' \neq i$ such that $c_{i'} > c_{i}$ we claim that $\payment{i'}{t^1} \leq \payment{i'}{t^{(i')}}$. In this case 
with $\ell' = \min\{j \in [m] \mid \prob{i'}{j} > 0 \}$,
the expected payment for action $i'$ under $t^1$ is 
\begin{align*}
\payment{i'}{t^1} = \sum_{j=1}^{m} \prob{i'}{j} \cdot t^1_j = \sum_{j=\ell'}^{m} \prob{i'}{j} \cdot \payment{i}{\tau} \leq \payment{i}{\tau} = \payment{i}{t^{(i')}}
\end{align*}
completing the proof.
\end{proof}

\section{Proof of Lemma~\ref{lem:linear-vs-welfare}}\label{app:linear-vs-welfare}

\begin{proof}[Proof of Lemma~\ref{lem:linear-vs-welfare}] 
If $W = 0$, then the linear contract $\langle (0,\ldots,0),1 \rangle$ is optimal and Equation~\eqref{eq:approx-bound} 
is immediate with $A = \{1\}$ and $\alpha_1 = 0$. So suppose $W > 0$.

Let $A \subseteq [n] \setminus \{1\}$ be a subset of actions formed by (i) excluding action 1 (the null action), (ii) excluding actions that cannot be implemented with a linear contract, and (iii) selecting a single action from each of the subsets (if any) of the remaining actions that have the same cost. 
Let $n' = |A|$. Note that since $W > 0$, it must be that $A \neq \emptyset$ and thus $1 \leq n' \leq n-1$. Relabel the actions $i \in A$ according to the smallest $\alpha \in [0,1]$ such that $\langle t, i \rangle = \langle (\alpha r_1, \ldots, \alpha r_m), i \rangle$ is incentive compatible, and let $\alpha_i$ denote the corresponding $\alpha$.  The sequences $\{R_i\}$, $\{W_i\}$, $\{c_i\}$ for $i \in A$ are strictly increasing in $i$, with $R_1 > 0$ and $c_1 \geq 0$.

Observe that for $i = 1$ we have $\alpha_1 = c_1/R_1$, while for $i > 1$ we have
\[
\alpha_i = \frac{c_i - c_{i-1}}{R_i - R_{i-1}}.
\]
Using this notation, we have
\[
\max_{\langle t,i \rangle \in \mathcal{L}(c,r',p)} U_P(\langle t,i \rangle) = \max_{i \in A} \; (1-\alpha_i)R_i.
\]

We next show an upper bound on the maximum social welfare. Note that
\[W = \max_{i \in A} W_i = W_{n'} = R_{n'} - c_{n'},
\]
where we used that the highest-welfare action is among the actions in $A$.
The upper bound follows from a lower bound on $(1-\alpha_i)R_i$, summed up over all $i \in A$.
 
For $i = 1$, we use that, by the definition of $\alpha_1$, it holds that 
$(1-\alpha_1) R_1 = (1-c_1/R_1)R_1 = R_1 - c_1$. For $i > 1$, we again use the formula for $\alpha_i$, to obtain that
\[
(1-\alpha_i) R_i = \left(1-\frac{c_i - c_{i-1}}{R_i - R_{i-1}}\right) R_i \geq (R_i - c_i) - (R_{i-1} - c_{i-1}),
\]
where we additionally used that $R_i/(R_i-R_{i-1}) \geq 1$.

Hence, by a telescoping sum argument,
\[
W = W_{n'} = (R_1-c_1) + \sum_{i =2}^{n'} \bigg((R_i-c_i)-(R_{i-1}-c_{i-1})\bigg) \leq \sum_{i=1}^{n'} (1-\alpha_i) R_i.
\]
Using this for the final inequality in the following, we thus obtain
\[
\max_{\langle t,i \rangle \in \mathcal{L}(c,r,p)} U_P(\langle t,i \rangle) = \max_{i \in A} \; (1-\alpha_i)R_i \geq \frac{1}{n'} \sum_{i=1}^{n'}(1-\alpha_i)R_i\ge \frac{1}{n'}W,
\]
giving the result.
\end{proof}

\section{Classic Contracts in Lower-Bound Instance}
\label{app:atmostone}

The following lemma shows that in the instance depicted in Figure~\ref{fig:gap-general}, the principal cannot obtain a utility greater than 1 using a classic contract.

\begin{lemma}
\label{lem:atmostone}
Let $n \geq 3$. Let $\gamma, \epsilon \in (0,1)$ and let $\delta = \epsilon \cdot \gamma^{n-2}$. Consider the parameterized instance $(c,r,p)$ with $n$ actions depicted in Figure~\ref{fig:gap-general}. Then 
\[
\max_{\langle t,i \rangle \in \mathcal{C}(c,r,p)} U_p(\langle t,i \rangle) \leq 1.
\]
\end{lemma}
\begin{proof}
We have $W_1 \leq W_2 \leq 1$, so there is nothing to show for these actions. Next consider any action $i$ such that $3 \leq i \leq n-1$. Suppose $\langle t,i \rangle$ implements action $i$. Since action $i$ puts zero probability on outcome $1$, we can without loss of generality assume that $t_1 = 0$. We derive a lower bound on $T_i(t)$, by considering only the IC constraint that compares action $i$ to action $i-1$. According to this constraint, we must have
\[
(1-\gamma^{n-i})t_2 + \gamma^{n-i}t_3 - c_i \geq (1-\gamma^{n-i+1})t_2 + \gamma^{n-i+1}t_3 - c_{i-1}.
\]
Since $(1-\gamma^{n-i})<(1-\gamma^{n-i+1})$ and $\gamma^{n-i}>\gamma^{n-i+1}$,  we obtain a lower bound on the expected payment $T_i(t)$ by setting $t_2 = 0$ and finding the smallest $t_3$ such that 
\[
\gamma^{n-i}t_3 - c_i \geq \gamma^{n-i+1}t_3 - c_{i-1}.
\]
Rearranging and substituting $c_i$ and $c_{i-1}$, this yields
\[
t_3 \geq \frac{1}{\gamma^{n-i}} \cdot \frac{c_i - c_{i-1}}{1-\gamma} = \frac{1}{\gamma^{n-i}} \left(\frac{1}{\gamma^{i-2}} - 1 \right).
\]
So the principal's utility from action $i$ is at most
\[
\gamma^{n-i} r_3 - \gamma^{n-i} t_3 \leq \frac{1}{\gamma^{i-2}} - \left(\frac{1}{\gamma^{i-2}} - 1 \right) = 1. 
\]

Next consider action $n$, and any IC contract $\langle t,n \rangle$. Since action $n$ puts zero probability on outcome $2$, we can without loss of generality assume $t_2 = 0$. To obtain an upper bound on the principal's utility, we proceed in a similar manner as before, except that now in addition to comparing to action $n-1$ we also compare to action~$1$.  
The comparison to action $1$ gives
\[
\delta t_1 + (1-\delta) t_3 - c_n \geq t_1,
\]
or equivalently
\[
t_3 - t_1 \geq \frac{c_n}{1-\delta}.
\]
An important consequence of this is that $t_3 - t_1 \geq 0$. Combining this with the fact that the agent does not want to deviate to action $n-1$, we obtain
\[
\gamma t_3 - c_{n-1} \leq \delta t_1 + (1-\delta) t_3 - c_n = 
t_3 - \delta(t_3-t_1) - c_{n} \leq t_3-c_n. 
\]
Rearranging and substituting $c_{n}$ and $c_{n-1}$, this yields
\[
t_3 \geq \frac{1}{1-\gamma} (c_n-c_{n-1}) = \frac{1}{\gamma^{n-2}} - 1.
\]
Hence the principal's utility from action $n$ 
is at most
\[
(1-\delta) r_3 - (1-\delta) t_3 \leq 1 - \delta,
\]
which completes the proof.
\end{proof}

\section{Algorithmic Implications}
\label{app:algorithms}

In this section, we show that the structural properties of optimal ambiguous contracts specified in Section~\ref{sec:computation} have important algorithmic implications. Specifically, they 
lead to polynomial-time algorithms for the optimal ambiguous contract problem.

\begin{theorem}[Computation]
\label{thm:computation}
There exists an algorithm that computes the optimal ambiguous contract in time $O(n^2 m)$. If the instance satisfies MLRP, then the running time improves to $O(n^2 + m)$.
\end{theorem}

\begin{proof}
We argue that for any given action $i \in [n]$ we can, in $O(nm)$ time, (1) decide whether it can be implemented by an ambiguous contract and (2) find the optimal contract $\langle \tau, i \rangle$ that implements it (if it is implementable). 
Applying this algorithm to all actions $i \in [n]$ and choosing the ambiguous contract 
that maximizes the principal's utility, we find the optimal ambiguous contract in $O(n^2m)$ time.

Fix any action $i \in [n]$. By Proposition~\ref{hadyn} action $i$ is implementable if there is no other action $i \neq i$ such that $p_{i'} = p_{i}$ and $c_{i'} < c_{i}$. For a given action $i' \neq i$ we can check whether $p_{i'} = p_i$ and $c_{i'} < c_i$ in $O(m)$ time. Applying this to all actions $i' \neq i$, we can test implementability in $O(nm)$ time.

For each action $i \in [n]$ that is implementable, we can determine an optimal ambiguous contract via Proposition~\ref{prop:opt-ambiguous}. Note that we can determine the set $A$ in $O(nm)$ time (simply by iterating over all actions $i' \neq i$ and checking whether $p_{i'} \neq p_{i}$). If $A = \emptyset$, then an optimal ambiguous contract is given by $\langle (0,\ldots,0), i \rangle$, and outputting this contract requires $O(1)$ time. On the other hand, if $A \neq \emptyset$, then for each of the at most $n-1$ actions in $A$ we can determine the maximum likelihood ratio outcome $j(i')$ in $O(m)$ time. We can then compute action $i'$'s contribution to $T$ in $O(1)$ time. We can thus compute the maximum likelihood ratio outcomes for all $i' \in A$ and $T$ in $O(nm)$ time. Outputting the optimal ambiguous contract described in Proposition~\ref{prop:opt-ambiguous} for this case takes $O(nm)$ time.
We conclude that we can determine an optimal ambiguous contract for action $i$ in $O(nm)$ time.

The improved running time for MLRP instances follows from the stronger characterization of optimal ambiguous contracts for such instances (Theorem~\ref{theo:unified MLRP SOP}). According to this characterization, to find the optimal ambiguous contract $\langle \tau, i \rangle$ that implements action $i$ (if one exists), we can first determine the indices $(\ell_i,h_i)$ (defined in bullet 2). Then, for each action $i' \neq i$ such that $c_{i'} \leq c_i$ we can find the minimum payment on outcome $h_i$ such that the agent prefers action $i$ over action $i'$; while for actions $i' \neq i$ such that $c_{i'} > c_i$ we can do the same with respect to outcome $\ell_i$ instead of $h_i$. Applying this procedure to all actions $i \in [n]$, we obtain an algorithm for finding the optimal ambiguous contract that runs in $O(n^2)$ time given access to the $(\ell_i,h_i)$ indices. The proof is completed by noting that for MLRP instances these indices are monotone (non-decreasing) in cost, and can thus be precomputed in  $O(n + m)$ time.    
\end{proof}

Notably, using similar ideas to the ones used in the proof of Theorem~\ref{thm:computation}, one can show that for monotone contracts, there exists an algorithm that computes the optimal ambiguous contract in time $O(n^2m)$. If the instance satisfies MLRP, then the running time improves to $O(n^2 + m)$.